\documentclass[12pt]{article}
\usepackage{subfigure}
\usepackage{a4}
\usepackage{epsf}
\usepackage{amssymb}
\usepackage{cite}
\newcommand{\be}{\begin{equation}}
\newcommand{\ee}{\end{equation}}
\newcommand{\bea}{\begin{eqnarray}}
\newcommand{\eea}{\end{eqnarray}}

\begin{document}
%%%%%%%%%%%%%%%%%%%%%%%%%%%%%%%
\def\C{{\mathbb{C}}}
\def\R{{\mathbb{R}}}
\def\s{{\mathbb{S}}}
\def\T{{\mathbb{T}}}
\def\Z{{\mathbb{Z}}}
\def\W{{\mathbb{W}}}
\def\Bbb{\mathbb}
\def\BZ{\Bbb Z} \def\BR{\Bbb R}
\def\BW{\Bbb W} 
\def\BM{\Bbb M} 
\def\e{\mbox{e}}
\def\BC{\Bbb C} \def\BP{\Bbb P}
\def\CP{\BC\BP}
%%%%%%%%%%%%%%%%%%%%%%%%%%%%
\begin{titlepage}
\vfill
\begin{center}
{\Large A Note on Dimer Models and D-brane Gauge Theories}\\[1cm]
Prarit Agarwal \footnote{E-mail: agarwalprarit@gmail.com},
P. Ramadevi\footnote{Email: ramadevi@phy.iitb.ac.in}\\
{\em Department of Physics, \\Indian Institute of Technology Bombay,\\
Mumbai 400 076, India\\[10pt]}
Tapobrata Sarkar\footnote{Email: tapo@iitk.ac.in}\\
{\em Department of Physics, \\ Indian Institute of Technology, 
\\ Kanpur 208016, India}\\
\end{center}
\vfill
\begin{abstract} 
The connection between quiver gauge theories and dimer models has been well
studied. It is known that the matter fields of the quiver gauge theories
can be represented using the perfect matchings of the corresponding dimer model.
We conjecture that a subset of perfect matchings associated
with an internal point in the toric diagram is sufficient to 
give  information about the charge matrix of the quiver gauge theory. 
Further, we perform explicit computations on some 
aspects of partial resolutions of toric singularities using dimer models. 
We analyse these with graph theory 
techniques, using the perfect matchings of orbifolds of the 
form $\BC^3/\Gamma$, where the orbifolding group $\Gamma$ may be noncyclic. 
Using these, we study the construction of the superpotential of gauge 
theories living on D-branes which probe these singularities, 
including the case where one or more adjoint 
fields are present upon partial resolution. Applying a combination
of open and closed string techniques to dimer models, we also study 
some aspects of their symmetries. 
\end{abstract}
\vfill
\end{titlepage}

\section{Introduction}\label{intro}

The recent spurt of interest in the study of D-brane gauge theories and
its relationship with dimer models in statistical mechanics arose after the
discovery of an infinite class of Sasaki-Einstein metrics with topology 
$S^2 \times S^3$. \cite{clpp},\cite{ms}. These spaces, which are defined
to be such that their metric cones are Ricci flat (and hence Calabi-Yau), 
arise in the extension of Maldacena's celebrated AdS/CFT duality 
(originally formulated in the context of $d = 4$, $N = 4$ supersymmetric 
Yang-Mills theory) to less supersymmetric $N = 1$ situations.  
It is well known that the low energy theory on a stack of D3-branes 
placed at the tip of such a Calabi-Yau cone has a gravity dual of the 
form $AdS_5 \times Y^5$, where $Y^5$ is Sasaki-Einstein. On the other
hand, the gauge theory living on the world volume of these D-brane can
be determined by using standard techniques pioneered in 
\cite{dm},\cite{dgm}.  It turns out that the toric 
description of the Sasaki-Einstein manifolds \cite{ms1} makes it 
possible to construct the full family of gauge theories dual to these 
spaces \cite{bfhms}. 

An important ingredient in the story is the role of brane tilings,  
which in turn leads us to the usage of the technology of dimer models 
in the description of D-brane gauge theories living on D-brane world
volumes. Dimer models, which have been well studied in areas of 
statistical mechanics and condensed matter physics (for reviews, 
see \cite{kast}, \cite{kenyon}) play a central role in much of this 
paper. The beautiful connection between dimer models and quiver gauge 
theories on D-brane probing orbifold singularities and their partial
resolutions was developed a few years back by Hanany 
and collaborators (for initial work in this direction, see \cite{han1},
\cite{fhkvw}. For comprehensive reviews, see \cite{kennaway},\cite{yamazaki}).  

In \cite{han1}, it was shown that there exists a connection between 
certain integers appearing in non-minimal resolutions of orbifold 
singularities, (as is typically seen by D-branes probing these), and 
combinatorial factors appearing in related dimer models. This provided 
an important computational tool in the study of D-brane gauge theories.
Dimer technology was then applied to a host of models and many aspects 
of the gauge theory living on D-brane world volumes have been understood 
from this perspective. Recently, in this context, various branches of
the vacuum moduli space of $N=1$ gauge theories have been comprehensively 
studied in \cite{masterspace}.  
On the other hand, in \cite{tapodimer}, a connection between dimer models
and closed string theories probing orbifold singularities was provided.
It was shown that dimers are naturally related to closed string theories
on orbifolds, via twisted sector R-charges of the latter.  

In this paper, we will discuss some issues relating to dimer models
and D-brane gauge theories, using both open and closed string perspectives
of dimers. We will see how these two descriptions nicely dovetail in the
context of non-compact orbifold theories, and using these we study some
aspects of symmetries of dimer graphs. We propose a conjecture that
a subset of perfect matchings corresponding to any internal
point of a toric diagram will be sufficient to study the faces
of the dimer diagram. Then, we write in an elegant way, the
charge matrix elements of a quiver gauge theories in terms
of this subset of perfect matchings.
Further, we study the construction
of gauge theories from dimer models via partial resolution of 
non-cyclic singularities, following the inverse algorithm of \cite{fhh},
and present some explicit calculations of the same  
verifying  our conjecture. We show how to obtain
the superpotential of certain partially resolved theories, using the
dimer description of the initial singularity, including the cases where
one or more adjoint fields might be present.   

The paper is organised as follows. In section 2, we review and recapitulate
certain known facts about dimer models as applied to orbifold gauge 
theories, before stating our conjecture that we discuss in the 
course of the paper. In section 3, we study cyclic abelian orbifolds of
$\BC^3$, combining certain ideas both from the open and closed string
pictures of the resolution of the same. In section 4, we will study
in detail the partial resolutions of some simple non cyclic orbifolds 
of the form $\BC^3/\Gamma$, concentrating on 
the cases where the orbifolding group $\Gamma$ is $\BZ_2\times\BZ_2$,
$\BZ_2\times\BZ_3$ and $\BZ_3\times\BZ_3$. We will also elaborate upon 
the role of adjoint fields that typically arise in the first two cases,
on partial resolution. Section 5 concludes with some discussions of 
our results.

\section{A Brief Review of Gauge Theories on Orbifolds and Dimers}

In this section, we will summarise and recapitulate the various ingredients
that we will need through the course of this paper. This section mostly
contains review material, and will serve to set the notations and  
conventions used in the rest of the paper. At the end, we also
specify a conjecture which will be verified and used in this paper.
To begin with, we will discuss the forward procedure (also called the
forward algorithm) \cite{dgm} that obtains the geometric data of a 
singularity from the quiver gauge theory of D-branes probing the same. 

Specifically, to deal with Abelian orbifold singularities, one conventionally 
uses a single D-brane probing the given singularity, extended in the transverse
directions and localised at the orbifold fixed point. Generically, such a 
D-brane (of type II string theory) is constructed \cite{dgm} 
by considering a theory of $r$ D-branes 
in $\BC^3$ and then projecting to $\BC^3/\Gamma$ 
where $\Gamma$ is the orbifolding group of rank $r$ 
that acts simultaneously on 
the space-time as well as the open string Chan Paton indices. The 
fields living on the D-brane are then the fields that survive the 
orbifolding action and the original gauge group $U(r)$ is broken to $U(1)^r$.  
We will be interested in the vacuum moduli space of this gauge theory.  

The gauge theory living on the D-brane world volume is characterised by two 
quantities - its matter content and its interactions. While the former
is captured by the D-terms in the gauge theory, the latter are 
described via the F-terms. For a single D-brane probing the orbifold 
singularity, the matter content consists of bi-fundamental fields, charged
under two $U(1)$ factors, and possible adjoints, which are uncharged under
any of the gauge groups. The bi-fundamental matter content 
is represented by a quiver diagram that gives the charge matrix 
$\Delta$ as its adjacency matrix, after the centre of mass $U(1)$ is
removed.  

Let us come to the F-term (superpotential) constraints.
Denoting the surviving fields of the gauge theory as $X_i, i=
1,\cdots,m$, it can be shown that the F-term equations are not all
independent, and that these can be solved in terms of $r+2$ parameters 
$v_j$,$j=1,\cdots,r+2$ (where $r$ is the rank of the orbifolding group) as
\begin{equation}
X_i = \prod_j v_j^{K_{ij}}
\label{revone}
\end{equation}
The matrix $K_{ij}$, $i = 1,\cdots,m$,$j=1,\cdots,r+2$ 
is the analogue of the matrix $\Delta$ for the F-terms. 

Conventionally, in toric descriptions of orbifold theories, 
having obtained the matrix $K$, we revert to its dual space, and 
solve for the dual matrix $T$, defined 
such that ${\vec K}.{\vec T} \geq 0$. $K$ being a 
$m \times (r+2)$ matrix, $T$ is typically of dimension $(r+2) \times c$, 
where $c$ is an integer that has to be determined on a case by case
basis. The dual matrix $T$ defines 
a new set of $c$ fields $p_{\alpha}, \alpha=1,\cdots c$. 
Determining the matrix $T$ is computationally intensive, but once 
it is obtained, the set of fields $v_i$ can be written in terms of 
the $p_{\alpha}$ as 
\begin{equation}
v_j = \prod_{\alpha}p_{\alpha}^{T_{j\alpha}}
\end{equation}
which, by eq. (\ref{revone}) implies 
\begin{equation}
X_i = \prod_{\alpha}p_{\alpha}^{\sum_j K_{ij}T_{j\alpha}}
\label{bifun}
\end{equation}

Now that we have a set of fields $p_{\alpha}$, we express
all physical variables in terms of these, and hence we need to find
the charges of these fields. Having written $r+2$ fields in terms of 
$c$ new fields, an extra $c - (r + 2)$ relations are needed to reduce
the extra variables to the original $r+2$. For this, we introduce a
new $U(1)^{c-r+2}$ gauge group, and gauge invariance conditions dictate
that the charges of the $p_{\alpha}$ fields are given by a 
matrix $Q$, which is the cokernel of $T$ and satisfies the relation
\begin{equation}
T.Q^t = 0
\end{equation}
Also, the charges of the $p_{\alpha}$ fields under the original 
$U(1)^r$ can be shown to be given by the matrix $VU$, where 
\begin{equation}
V.K^t = \Delta,~~~~U.T^t = I
\end{equation}
Note that since the matrix $VU$ encodes the information of the charges
of the new variables $p_{\alpha}$ in terms of the original set of $U(1)$s,
they naturally denote the D-term constraints in terms of the new 
fields (and hence has, associated to each, a Fayet-Illiopoulos (FI) parameter), 
whereas the matrix $Q$ carries information about the 
redundancies in the parametrization of the new variables. It is thus
natural to label these matrices as $Q \equiv Q_F$ and $VU \equiv Q_D$.
Now, concatenating $Q_F$ and $Q_D$, the kernel of the resulting matrix
gives the toric data of the singularity that is being probed.  
In summary, then, the above prescription gives us a holomorphic quotient
description of the toric variety. As an example, for the $\BC^3/\BZ_3
\times\BZ_3$ singularity \cite{bglp}, the space of F-flatness 
conditions is described 
as the holomorphic quotient $\BC^{42}/\left(\BC^*\right)^{31}$ and the
moduli space of vacua is obtained by acting on this (with certain point
sets removed, as dictated by the choice of FI parameters) the complexification
of the original gauge group $U(1)^8$. 

The symplectic description of the above singularity can be constructed
using the procedure due to \cite{mp},\cite{bglp}. To illustrate
this, we will again consider the singularity $\BC^3/\BZ_3
\times\BZ_3$. Here, one begins with the closed string twisted sectors,
and inserts fractional points in the $\BZ^{\oplus 3}$ lattice corresponding
to the closed string R charges. Restoring integrality in the lattice then
gives the toric data for the resolution of the orbifold. In this particular
case, there are seven internal points that need to be added, and the 
symplectic description is a quotient of $\BC^{10}$, after 
removing a certain point set, by a $U(1)^7$ action \cite{bglp}. The
map between the FI parameters of the D-brane gauge theory to the 
FI parameters in the closed string description can be computed, and 
determines the physicality of the gauge theory living on the D-brane upon
partial resolution. Note that the closed string and the D-brane gauge 
theory description of the geometry of the singularity are at different 
points in its K\"ahler moduli space. Whereas the former describes the
geometry at the orbifold point, the latter provides a description of 
the geometry at the conifold point. There are many important differences
between the two descriptions, e.g. the open string theory does not 
probe the non-geometric phases of the theory \cite{dgm}.  
Importantly, the open string description is typically non-minimal, in
the sense that the points in the toric diagrams appear with multiplicities.
These multiplicities have been studied extensively in the last few years,
particularly by appealing to the inverse algorithm developed in \cite{fhh},
and it has been realised that they can be used to construct different 
gauge theories that flow to the same universality class in the infrared. 
This is called toric duality, which can be shown to be equivalent to
Seiberg duality of gauge theories.  

In \cite{han1}, it was realised that the description of D-brane gauge
theories has a striking correspondence with brane tilings and its 
underlying dimer models, the latter having been well studied in the 
context of statistical mechanics. \footnote{Physically, brane tilings 
represent a collection of NS5 and D5 branes. Each edge in a perfect 
matching of the brane tiling (to be discussed momentarily) 
is referred to as a dimer. We will refer to 
dimer models and brane tilings in the same spirit, and the distinction 
should be obvious to the reader from the context.} Dimer models refer to 
the statistical mechanics of bipartite graphs, which consist of a possibly 
infinite number of vertices, with the property that each vertex
can be colored black or white, with no two vertices of the same color being 
adjacent (in the sense of the nearest neighbor). Given such a graph, one
can define two concepts : its fundamental domain and perfect
matchings. The fundamental domain of a bipartite graph is essentially its
unit cell. Perfect matchings of the graph consist of a subset of 
edges (called dimers, since they connect two vertices of the graph) such that
each edge connects one black to one white vertex. In the context of string 
theory, these graphs appear to be naturally related to orbifold theories
and their resolutions,
and for these, the fundamental domain can be obtained by extending that 
for the flat space case. For the purpose of this paper, we will be mostly
interested in $N=1$ gauge theories, i.e orbifolds of $\BC^3$. 
Non-orbifold theories can be obtained as partial resolutions of these, or in 
some cases by adding ``impurities'' to the orbifold theories \cite{fhkvw}.
Dimer models provide the right variables for the study of D-brane gauge
theories that probe Calabi Yau singularities, and it was realised in
\cite{han1} that the connection between the two arise via the properties 
of the Kasteleyn matrix used to characterise the former. \footnote{For 
a review, the reader is referred to \cite{kennaway}.} Broadly speaking,
one can translate between objects in the dimer model and those in 
the gauge theory using the following dictionary : faces, nodes and edges
in the dimer model correspond to the gauge groups, superpotential terms
and bifundamental (or adjoint) fields in the gauge theory. We will discuss
these in details in the next part of the paper, but before we move on, 
let us illustrate the concept of the matching matrix which will be very
useful for us later. Given a dimer model, a perfect matching 
represents a collection of bifundamental (and possibly adjoint) fields, and is 
a subset of the full set of fields in the gauge theory. Given a set of
perfect matchings $\{p_{\alpha}\}$, we can 
define the matching matrix as
\begin{equation}
{\cal M}_{i\alpha} = \langle e_i,p_{\alpha}\rangle
\label{mm}
\end{equation}
where ${\cal M}_{i\alpha}$ represents a Kronecker delta function in the sense
that it takes value $1$ if the bifundamental field represented by 
the edge $e_i$ is contained 
in the matching $\alpha$, and vanishes otherwise. Since there is 
a one to once correspondence between perfect matchings in the dimer
model and GLSM fields in the corresponding orbifold theory \cite{francovegh}, 
in terms of the matching matrix, eq. (\ref{bifun}) can be written as 
\begin{equation}
X_i = \prod_{\alpha=1}^cp_{\alpha}^{{\cal M}_{i\alpha}}
\end{equation} 
In addition, it can be shown that the redundancy matrix corresponding to
the matching matrix ${\cal M}$ gives us the F-term charges in the D-brane gauge 
theory. A further concept that we will need is that of face symmetries of
a given dimer model. Given two perfect matchings $p_1$ and $p_2$ of a dimer
model, their difference gives a collection of closed curves in the dimer 
graph. A closed curve that goes around a face of the graph is related to 
a face symmetry of the model. As we have mentioned, faces in the dimer
model correspond to gauge groups in the D-brane gauge theory. Hence, the
face symmetries are related to the D-terms in the latter. 
We will use these facts extensively in the next couple of sections. 

Before we end this section, let us briefly point out an alternative
way of looking at dimer models, i.e from closed string theory.  
In \cite{tapodimer}, it was shown that dimer models are related
to closed string theories on non-compact orbifolds of $\BC^3$ (and
also of $\BC^2$), via the closed string twisted sector R-charges, which,
in a sense, are analogues of the height functions \cite{kenyon}. In
particular, it was shown that perfect matchings in dimer models can 
be interpreted as twisted sector states, via the assignment of certain
fractional weights to the edges of the dimer (that depends on the 
particular orbifold theory being considered). 
This also serves to specify the position
of a given perfect matching in the toric diagram. It was further observed 
in \cite{tapodimer} that a given state with a certain assignment of R-charges 
correspond to more than one perfect matching in the dimer model. These 
are in one to one correspondence with the multiplicities of these
states in the open string picture of probe D-branes, although, as we
have said, closed strings and D-branes probe these orbifold singularities 
in different ways \cite{dgm}.  

Having reviewed the basic setup, we now proceed to the main part of the paper.
In this paper, we will perform some explicit computations using the concepts 
mentioned above. In particular, apart from the cyclic orbifolds of the
form $\BC^3/\BZ_n$, we will use dimer model techniques to study, in details, 
the partial resolution of the orbifolds $\BC^3/\BZ_2\times\BZ_3$ and 
$\BC^3/\BZ_3\times\BZ_3$. Let us highlight some of the issues that we will 
make precise in the rest of the paper, and which will be needed to study
the partial resolutions of non-cyclic orbifolds. We state them in the 
form of a conjecture and will provide evidence for these in what follows.

\noindent
{\bf Conjecture:} \\
The face symmetries for dimer models which correspond to toric singularities 
can be written entirely in terms of those perfect matchings that correspond 
to the internal points of the toric diagram, whenever these are present.

\noindent 
We can elaborate this conjecture in an algebraic way as follows:
suppose $\{p^a_{\alpha}\}$ is the set of perfect matchings associated
with an internal point in the toric diagram such that the 
closed contour formed by them goes around the
$a$th face of the dimer model (in clockwise orientation). Let 
$F_a$ denote the combination of the perfect matchings that form the above
contour, i.e 
\begin{equation}
F_a = \sum_{\alpha}{\rm sign}\left(\alpha, a\right)p^a_{\alpha}
\end{equation}
where ${\rm sign}\left(\alpha, a\right) = \pm 1$ if the edge contributed by
$p_{\alpha}$ is traversed from the white to black (resp. black to white) node.

We can now write the elements of the charge matrix of the matter 
field $X_i$ in the quiver gauge theory as 
\begin{equation}
d_{a i}=\langle e_i,F_a\rangle
\label{conj1}
\end{equation}
where, as before, $e_i$ is the edge denoting the bifundamental field $X_i$.
From the results of \cite{kennaway}, it is easy to see that 
eqn. (\ref{conj1}) is true by rewriting the equation
in two steps. We construct the matrix $A$ whose elements are
\begin{equation}
A_{a \alpha} = \langle F_a, p_{\alpha}\rangle
\end{equation}
Then in terms of $A$ and the matching matrix ${\cal M}$, we can 
write the quiver charge matrix $d$ as
\begin{equation}
d= A {\cal M}^t
\end{equation}
This implies that $A= Q_D$, which can be checked from its definition. 

In the next section, we check this conjecture in a few simple orbifold
setting. Later on, we will discuss how this is verified for more
complicated orbifold as well as non orbifold singularities.
                                                                                
\section{Cyclic orbifolds of $\BC^3$}

In this section, we study some properties of cyclic orbifolds of $\BC^3$,
from a dimer model perspective.
In particular, we will focus on the simple examples of $\BC^3/\BZ_3$ and
$\BC^3/\BZ_5$. These have one and two interior points, respectively in their
toric diagram. 
\subsection{The orbifolds $\BC^3/\BZ_3$ and $\BC^3/\BZ_5$}

\begin{figure}%[h]
\centering
\subfigure[Dimer Diagram]
{\label {dimerz3}
\epsfxsize=2.45in
\hspace*{0in}\vspace*{.2in}
\epsffile{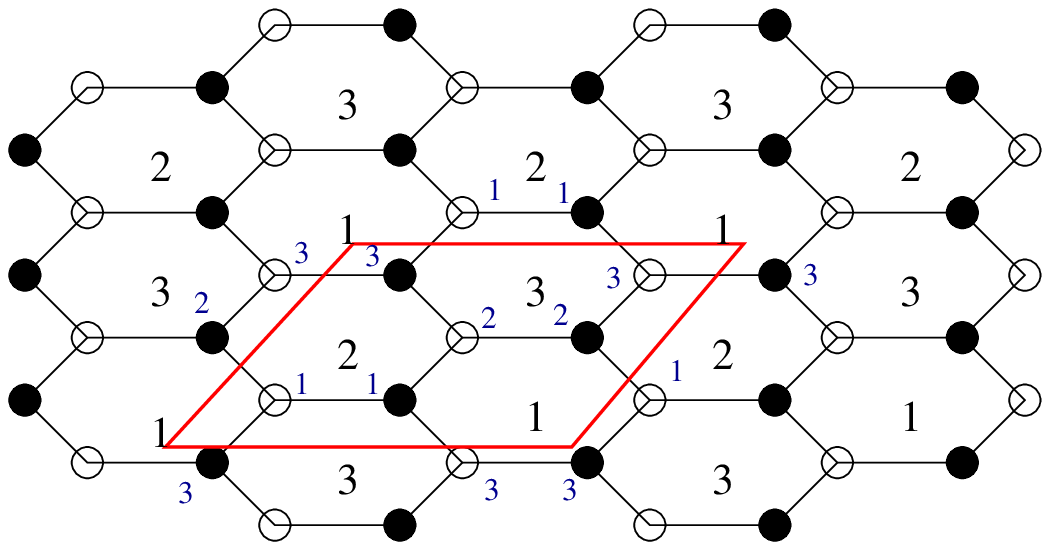}}
%\caption{\small Dimer model and perfect matchings for the orbifold
%$\BC^3/\BZ_3$} 
%\label{c3z3}
\hspace{.25in}
\subfigure[Perfect matchings ]
{\label{pmz3}
\epsfxsize=2.45in
\hspace*{0in}\vspace*{.2in}
\epsffile{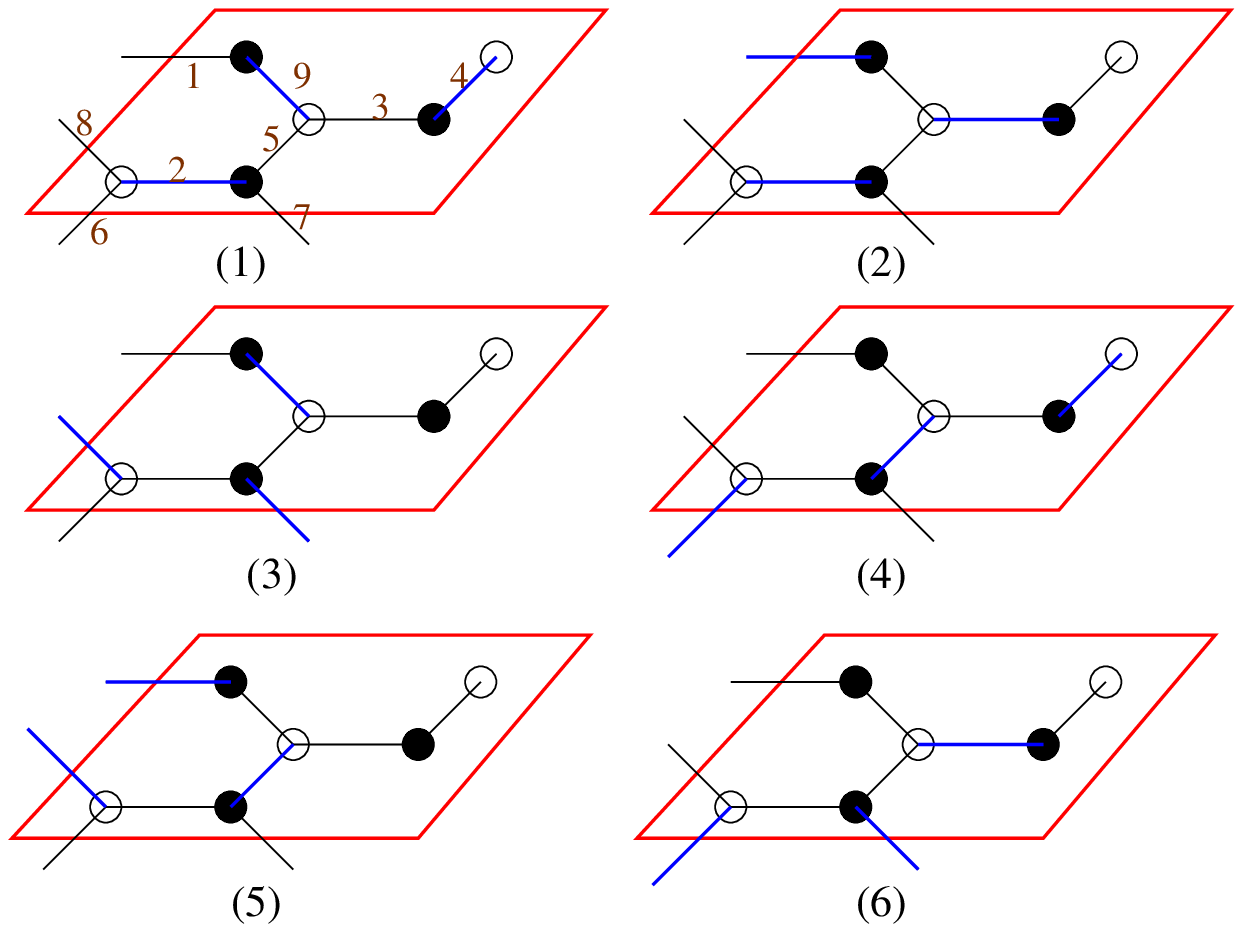}}
\caption{Dimer model and perfect matchings for the orbifold $\BC^3/\BZ_3$} 
\label{c3z3}
\end{figure}

Let us begin with the orbifold $\BC^3/\BZ_3$, which is also 
the cone over the zeroth del Pezzo surface. From the closed string perspective,
the toric diagram is obtained by restoring integrality in the 
$\BZ^{\oplus 3}$ lattice consisting of the points $\left(1,0,0\right)$,
$\left(0,1,0\right)$, $\left(0,0,1\right)$ and $\left(\frac{1}{3},
\frac{1}{3},\frac{1}{3}\right)$, where the fractional point corresponds to
the only marginal twisted sector in the theory, (the other one being
irrelevant). The combinatorics of this model can be obtained by weighing
the the three distinct edges of the dimer model shown in 
fig. \ref{dimerz3} 
by the vectors $\left(\frac{1}{3},0,0\right)$, 
$\left(0,\frac{1}{3},0\right)$ and $\left(0,0,\frac{1}{3}\right)$. In 
fig. \ref{pmz3}, the perfect matchings numbered $2$, $3$ and $4$ have
weights $\left(1,0,0\right)$, $\left(0,1,0\right)$ and $\left(0,0,1\right)$
respectively. The matchings numbered $1$, $5$ and $6$ have weights
$\left(\frac{1}{3},\frac{1}{3},\frac{1}{3}\right)$, and these correspond
to the marginal twisted sector of the theory. In fact, these
three perfect matching represent the multiplicity of the internal
point in the toric diagram. Hereafter, we call the perfect matchings 
associated with an internal point as {\bf internal perfect matchings}.
%%%%%%%%%%%%%%%%%%%%%%%%%%%%
%%%%%%%%%%%%%%%%%%%%%%%%%%%%
The Kasteleyn matrix \cite {kennaway} will be
\begin{eqnarray}
K(Z,W) = 
\begin{pmatrix}{
a11&a12 W&a13 Z\cr
a21&a22&a23\cr
{a31 \over Z}&a32&{a33 \over W}
}
\end{pmatrix}
\end{eqnarray}
where $aij$ is the label which keeps track of the edge connecting
the $i$-th white node to the $j$-th black node and the nodes
have been numbered in fig. \ref{dimerz3}. The
determinant of $K(W,Z)$ will give six terms corresponding to the 
six perfect matchings:
{\small
\begin{eqnarray}
det~ K&=& 
-a13a22a31+a12a23a31{W \over Z}-a11a23a32-a12a21a33+a13a21a32 Z+
\nonumber\\
~&~&a11a22a33 {1 \over W}~.
\end{eqnarray}}
The last term, for example, is an algebraic representation of the 
second perfect matching
shown in fig. \ref{pmz3}. In  non-trivial toric Calabi-Yau geometries 
where the number of perfect matchings is large, the algebraic way 
of representing perfect matchings makes it easier 
to determine face symmetries. 

The matching matrix (\ref{mm}) is 
{\small 
\begin{eqnarray}
{\cal M}=\begin{pmatrix}
{ & p_1 & p_2 & p_3 & p_4 & p_5 & p_6\cr
X_1 & 0 & 1 & 0 & 0 & 1 & 0\cr
X_2 & 1 & 1 & 0 & 0 & 0 & 0\cr
X_3 & 0 & 1 & 0 & 0 & 0 & 1\cr
X_4 & 1 & 0 & 0 & 1 & 0 & 0\cr
X_5 & 0 & 0 & 0 & 1 & 1 & 0\cr
X_6 & 0 & 0 & 0 & 1 & 0 & 1\cr
X_7 & 0 & 0 & 1 & 0 & 0 & 1\cr
X_8 & 0 & 0 & 1 & 0 & 1 & 0\cr
X_9 & 1 & 0 & 1 & 0 & 0 & 0 }
 \end{pmatrix}
\end{eqnarray}}
and the charge matrix $Q_F$ can be calculated to be
\begin{equation}
Q_F = \left(1,-1,-1,-1,1,1\right)
\label{c3z3Fcharge}
\end{equation}
so that the masterspace \cite{masterspace} for this theory 
is the space $\BC^6$ modded out by a $U(1)$
with the above charges. Let us now check the two conjectures mentioned in 
the last section, for this example. First, we discuss the second conjecture
regarding the face symmetries. From the closed string point of view, 
our main observation is that any symmetry associated with the dimer 
covering should necessarily involve combinations of perfect matchings 
which force the total closed string R-charge to zero. A symmetry associated 
with a particular face in the dimer covering is thus equivalent to the 
closed string R-charges vanishing around that face. This can be seen 
from fig.\ref{dimerz3}. We look for a minimum number of perfect
matchings whose combination will enclose the face. Interestingly, 
the face symmetries are most easily obtained 
by taking pairwise differences of the internal perfect matchings 
(corresponding to the twisted sector charge 
$\left(\frac{1}{3},\frac{1}{3},\frac{1}{3}\right)$). 
These faces $F_a$'s are given by the combinations 
$F_1=p_5-p_6$, $F_2= p_1 - p_5$,  and $F_3=p_6 - p_1$, where the 
subscript on $p$ refers to the matching numbers as in 
fig. (\ref{c3z3}). \footnote{In this case, we observe that the redundancy in
the matching matrix is the symmetry which involves the external matchings
as well. The only such combination will involve the charge matrix in eq.
(\ref{c3z3Fcharge}).} Given the labeling of the edges in fig. (\ref{c3z3}),
conjecture 1 of the last section can now be easily shown to yield 
the quiver charge matrix for the orbifold $\BC^3/\BZ_3$
{\small
\begin{eqnarray}
d=
\pmatrix{ &X_1 &X_2&X_3&X_4&X_5&X_6&X_7&X_8&X_9\cr
F_1&1&0&-1&0&1&-1&-1&1&0\cr
F_2&-1&1&0&1&-1&0&0&-1&1\cr
F_3&0&-1&1&-1&0&1&1&0&-1\cr
}\label{c3z3d}\end{eqnarray}}
  
\begin{figure}
\centering
\epsfxsize=2.2in
\hspace*{0in}\vspace*{0.2in}
\epsffile{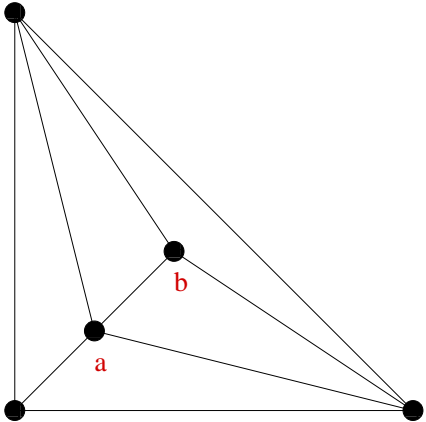}
\caption{\small Toric diagram for the supersymmetric
orbifold $\BC^3/\BZ_5$.} 
\label{c3z5toric}
\end{figure}

\begin{figure}
\centering
\epsfxsize=3.2in
\hspace*{0in}\vspace*{0.2in}
\epsffile{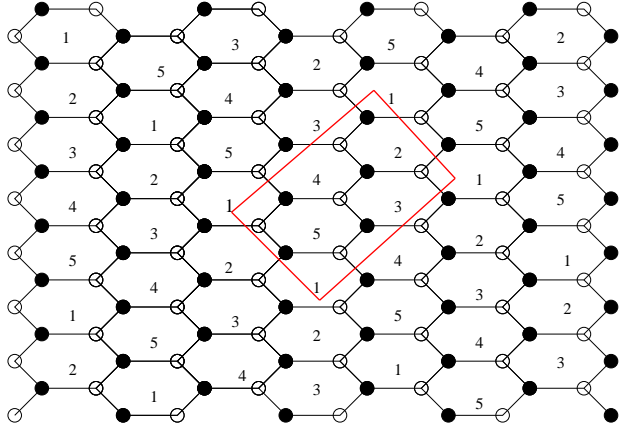}
\caption{\small The dimer model for the supersymmetric
orbifold $\BC^3/\BZ_5$, with the fundamental cell is shown in red.}
\label{c3z5}
\end{figure}

We now turn to the orbifold $\BC^3/\BZ_5$, with the orbifolding action
being 
\begin{equation}
\left(Z^1,Z^2,Z^3\right)\to\left(\omega Z^1,\omega Z^2,\omega^3 Z^3\right) 
\end{equation}
where $\omega = e^{\frac{2\pi i}{5}}$. This is the simplest case where
there are two internal points in the toric diagram. These correspond to the two
marginal twisted sectors in the closed string theory, with twisted sector
R-charges $\left(\frac{1}{5},\frac{1}{5},\frac{3}{5}\right)$ and 
$\left(\frac{2}{5},\frac{2}{5},\frac{1}{5}\right)$
\cite {tapodimer}. Inserting these points in the $\BZ^{\oplus 3}$ 
lattice along with the unit vectors, the toric 
diagram is obtained as shown in fig. (\ref{c3z5toric}).
Note that the two internal points marked $a$ and $b$ in Fig. 2 
are with multiplicity $5$ and $5$. 
The dimer model for this orbifold is shown in fig. (\ref{c3z5}).   
%%%%%%%%%%%%%%%%%%%%%%%%%%%%
%%%%%%%%%%%%%%%%%%%%%%%%%%%%%
In the appendix, for completeness, we have listed the $10$ internal perfect 
matchings for 
the orbifold $\BC^3/\BZ_5$ \cite{tapodimer} in fig. (\ref{app0}). 
In this case, there are two types of
perfect matchings (corresponding to the two marginal twisted sectors mentioned 
above). Looking for a minimum number of perfect matching enclosing a face,
we confine to pair wise differences. It is not difficult to see that
such face symmetries will involve either
of the two sets of internal perfect matchings. Also, the fact that the twisted
sector  $R$-charge vanishes along a face reinforces that there is 
no mixing between the two sets of internal perfect matchings. 

In this example, from fig. (\ref{app0}), it can be seen  that the 
face symmetries can be constructed
either by the difference in matchings $p_1-p_7$, $p_2-p_4$, $p_7-p_5$ and
$p_5-p_2$, or, equivalently, from the matchings $p_{10}-p_3$, $p_8-p_9$,
$p_3-p_8$ and $p_6-p_{10}$. Both these choices can be seen to give 
rise to the same quiver charges. The same analysis goes through 
for orbifold toric diagrams with 
multiple interior points, and the face symmetries can be written down with 
perfect matchings corresponding to a single marginal twisted sector. Given that
for orbifolds of the form $\BC^3/\BZ_n$, twisted sectors appear as internal 
points in the toric diagram, 
\footnote{This is not necessarily true for orbifolds
with non-isolated singularities, which might have points on the external edges
of the toric diagram. We will restrict our analysis to orbifolds theories
which have isolated singularities only. This means that we choose orbifolds
of the form $\BC^3/\BZ_n$ with $n$ a prime number, and the orbifolding action
$\left(Z^1,Z^2,Z^3\right) \to \left(Z^1,\omega^p Z^2,\omega^q Z^3\right)$
where $1+p+q=0 {\rm mod} n$ is such that $p$ and $q$ are relatively prime to
$n$. However, for non-isolated singularities, one could always treat the points
on the edges of the toric diagram as internal points, and our analysis can be
easily extended to these cases.} this means that for generic orbifold theories, 
the face symmetries are given by combinations of internal points only. We will
see in the next section that this is true for non-cyclic orbifolds as well,
and we conjecture that this is also true for non-orbifold theories with internal
points.   
%%%%%%%%%%%%%%%%%%%%%%%%%%%%
%%%%%%%%%%%%%%%%%%%%%%%%%%%%%

Before we end this section, let us briefly point out another interesting 
aspect of dimer model combinatorics as applied to cyclic orbifolds,  
using the closed string approach. 

\subsection{Exploring different regions in K\"ahler moduli space}

\begin{figure}
\centering
\epsfxsize=3.2in
\hspace*{0in}\vspace*{0.2in}
\epsffile{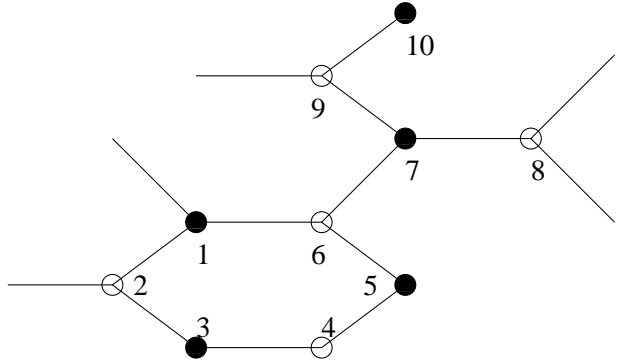}
\caption{\small The fundamental region for the dimer model for the 
orbifold $\BC^3/\BZ_5$, redrawn from fig. (\ref{c3z5})}
\label{c3z5fund}
\end{figure}

Let us first consider the orbifold $\BC^3/\BZ_5$ as an example (we
momentarily generalise the results to generic $\BZ_n$ orbifold theories). 
The orbifolding action on the coordinates is 
\begin{equation}
\left(Z^1,Z^2,Z^3\right) \to \left(\omega Z^1,\omega Z^2,\omega^3Z^3\right)
\end{equation}
where $\omega = e^{\frac{2\pi i}{5}}$.
$\BC^3/\BZ_5$ has a closed string $U(1)$ GLSM description in terms of four
fields $\phi_i, i=1,\cdots,4$ with $U(1)$ charges 
\begin{equation}
Q = \left(1,1,3,-5\right)
\end{equation}
There is a single D-term constraint in the theory,
\begin{equation}
|\phi_1|^2 + |\phi_2|^2 + 3|\phi_3|^2 - 5|\phi_4|^2 + r = 0
\end{equation}
where for $r \gg 0$, the field $\phi_4$ acquires a large positive value, which
breaks the $U(1)$ symmetry into a $\BZ_5$, and the massless fields $\phi_i, 
i=1,2,3$ transform according to the unbroken $\BZ_5$ symmetry. In the opposite
limit $r \ll 0$, the theory is given by a line bundle over a suitable weighted 
projective space \cite{wittenphases}. At other points in K\"ahler 
moduli space, there might also be a local orbifold singularity, where the
orbifolding action involves the discrete group $\BZ_3$, or the theory
might look like flat space. 
These can be seen by splitting the toric diagram of fig. (\ref{c3z5toric})
from the point denoted as {\bf a}. This gives rise to three 
triangles, two of which do not contain an internal point (and hence represent
the flat space $\BC^3$) and the third one has one internal point, marked
{\bf b} in fig. (\ref{c3z5toric}) and is identified with the orbifold
$\BC^3/\BZ_3$. We can equivalently split the toric diagram from point 
{\bf b}, and this can be interpreted as the case corresponding 
to the orbifolding action
$\left(Z^1,Z^2,Z^3\right) \to \left(Z^1,\omega^2Z^2,\omega^2Z^3\right)$,    
where $\omega$ is the fifth root of unity. We will focus on the first
case here. \footnote{This can also be 
visualised by providing suitable vevs to fields $\phi_1$,
$\phi_2$ and $\phi_3$, and studying the sigma model metrics that result from
the same. Whereas in the first two cases, we recover flat space, the third
can be seen to result in the orbifold $\BC^3/\BZ_3$. 
These theories are infinitely 
separated in space.} It is possible to understand this from a dimer
model perspective. Let us see if we can substantiate this. Consider the 
fundamental region for the dimer covering of the orbifold $\BC^3/\BZ_5$ redrawn 
from fig. (\ref{c3z5}) in fig. (\ref{c3z5fund}) 
%%%%%%%%%%%%%%%%%%%%%%%%%%%%
%%%%%%%%%%%%%%%%%%%%%%%%%%%%%
Purely from a combinatorial viewpoint (distinct from Higgsing the theory
as in \cite{uranga2}), 
note that the fundamental region in
fig. (\ref{c3z5fund}) can be thought of as a gluing of three separate pieces
which also qualify as fundamental regions of orbifolds of lower rank - namely,
the lower hexagon consisting of the nodes $1$ to $6$, with three external lines,
i.e a total of $9$ bifundamentals, the points $7$ and $8$, and the 
points $9$ and $10$, with both the latter ones having three bifundamental 
fields each associated to them. However, now the 
leg from node $6$ has to be identified with node $3$. The latter process    
can be thought of as the analogue of normalisation of $U(1)$ charges upon reduction of 
the orbifolding group \cite{tapo2}. This can be generalised as follows. A cyclic
orbifold $\BC^3/\BZ_n$ will contain, in the fundamental domain of its perfect 
matching, $2n$ nodes corresponding to the $2n$ terms in the superpotential, and 
$3n$ bifundamental fields. In order to construct other 
locally orbifold theories   
at different points in the K\"ahler moduli space, we split the original domain
into three parts at a given node (for instance node 7 for $\BZ_5$
as shown in fig. (\ref {c3z5fund})), 
with each of these parts carrying one of the
three edges associated to that node, with the constraint that the three 
resulting parts have an even number of nodes and an odd number of edges. The 
latter constraint is required to make the resulting theories $\BC^3$
orbifolds (or flat space). This generically gives us the theories corresponding
to splitting the toric diagram along one of its internal points.
This procedure can be seen to go through for higher rank 
orbifolding groups as well, which might have more than one 
distinct orbifolding action. 

Let us summarise our discussion so far. We have studied
the supersymmetric orbifolds of the form $\BC^3/\BZ_n$ using a combination of
open string and closed string methods. 
We saw that any symmetry associated to the dimer model implies the 
vanishing of the total closed string R-charge (a subset of which
gives the face symmetries). 
For toric (cyclic) orbifolds with more than one 
internal point, these symmetries can be constructed out of any given 
type of internal point corresponding to a particular R-charge. We will
see in the next section that this result is valid for non-cyclic 
orbifolds as well. Further, we expect that for non-orbifold
theories also, the face symmetries can be constructed out of internal
points only, whenever these are present.  
We have also seen that data 
about the (local) orbifolds of lower rank corresponding 
to different points in the K\"ahler moduli space of the original theory are
encoded in the dimer covering of the higher rank theory, and can be analysed by 
splitting (or equivalently gluing) sub diagrams along nodes. These are the
main results of this section.  We now study some aspects of non cyclic 
orbifolds of $\BC^3$.
 
\section{Non cyclic orbifolds of $\BC^3$}

In this section, we will study some simple non-cyclic orbifolds of $\BC^3$, whose
partial resolutions generically give non-orbifold theories. 
From the point of view of dimer coverings, partial resolutions can be obtained
by removing edges from the dimer diagram, which corresponds to a Higgsing 
process in the D-brane gauge theory \cite{han1}. 
Our aim in this section will be to study these in some details. As is
known, arbitrary removal of edges from a dimer model may not correspond 
to physical D-brane gauge theories. First of all, we note that from the 
field theory point of view, the Higgsing procedure will not be meaningful if we
remove two adjacent edges from a dimer diagram (i.e edges that meet at a single
node). Hence, we can eliminate this possibility by giving vevs to edges that
do not meet at a node. Even then, the theory is not guaranteed to be consistent,
as we will see. An easy way to check consistency of the gauge theory is to
derive the superpotential of the resulting theory after Higgsing. A 
consistent superpotential is one in which the open string modes appear 
exactly twice, and as we will see in the next few subsections, even after 
removing non-adjacent edges from a dimer covering, the resulting theory might 
have an inconsistent superpotential. Unfortunately, there is no general 
prescription to a priori determine the set of physical gauge theories that might
arise due to Higgsing, and one has to proceed on a case to case basis.  

In discussing partial resolutions of abelian orbifolds, we will use 
directly the matching matrix for the ``parent'' theory (which will 
give us the various partially resolved ``daughter'' theories). Indeed the
Higgsing procedure can be simply implemented by starting with the full
matching matrix ${\cal M}$ (whose rows we label by the fields on the probe
D-brane world volume and the columns are the perfect matchings in which
these fields occur), and then directly removing those rows which 
correspond to fields that acquire vevs, and the columns that have non-zero
entries corresponding to these rows. This will give us the reduced matching
matrix for the partially resolved singularity, from which the gauge theory
data can be read off by a prescription similar to the inverse algorithm
due to \cite{fhh}. Namely, following the notation conventions of section
(2), given the reduced matching matrix $M_r$, we calculate the redundancy 
matrix $Q_{F(r)}$, whose kernel gives us the reduced $T$ matrix, which we
label by $T_r$. The dual of the matrix $T_r$  is the reduced $K$ matrix 
(of section 2) inherited by the (partially resolved) daughter
singularity. This now can be integrated to give the superpotential of 
the reduced (non-orbifold) theory. In order to avoid notational 
complications, we will denote the reduced $K$ matrix of the daughter
singularities by the symbol ${\cal K}_r$. It will be understood that 
the reduced matching matrices for these are obtained from 
${\cal M}_r = {\cal K}_r^t . T_r$. In this procedure,
one can work directly in terms of the fields in the D-brane gauge theory,
and the Higgsing procedure, whenever physical in the sense of the
previous paragraph, is guaranteed to give us a resulting physical gauge theory. 

The difficulty with the above procedure seems to be that it is incapable
of handling adjoint fields, and these have to be added by hand
in order to obtain a consistent superpotential \cite{fhh}. In the
field theory, this can be understood by looking at the transformation
properties of the massless modes after the Higgsing procedure;  
in the dimer model description, the equivalent statement is that 
in the matching matrix, the number of non-zero field
entries in any perfect matching is always the same. E.g, for orbifolds, 
the total number of edges participating in a perfect matching is always 
$n$, where $n$ is the rank of the orbifolding group. It is easy to see
that the same statement goes over for non-orbifold theories as well. 
This will be useful for us in what follows in order to describe theories
that give rise to one or more adjoint fields upon Higgsing.
 
\subsection{The $\BC^3/\BZ_2\times\BZ_2$ singularity}

\begin{figure}[h]
\centering
\epsfxsize=4.5in
\hspace*{0in}\vspace*{.2in}
\epsffile{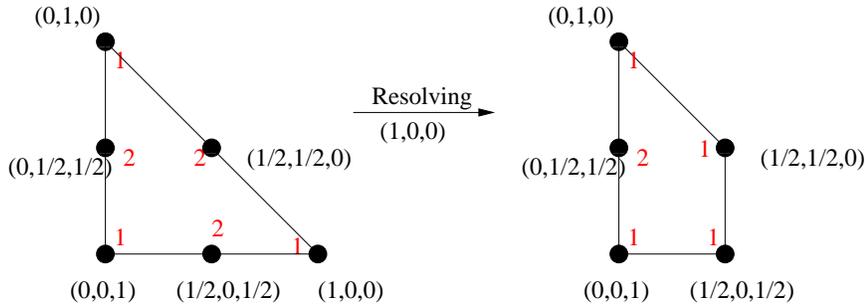}
\caption{\small Toric diagram for the partial resolution of the singularity 
$\BC^3/\BZ_2\times\BZ_2$ to the SPP singularity. We have also shown the
closed string R-charges of the parent orbifold theory in both the diagrams.}
\label{fig1}
\end{figure}

\begin{figure}[h]
\centering
\epsfxsize=6.0in
\hspace*{0in}\vspace*{.2in}
\epsffile{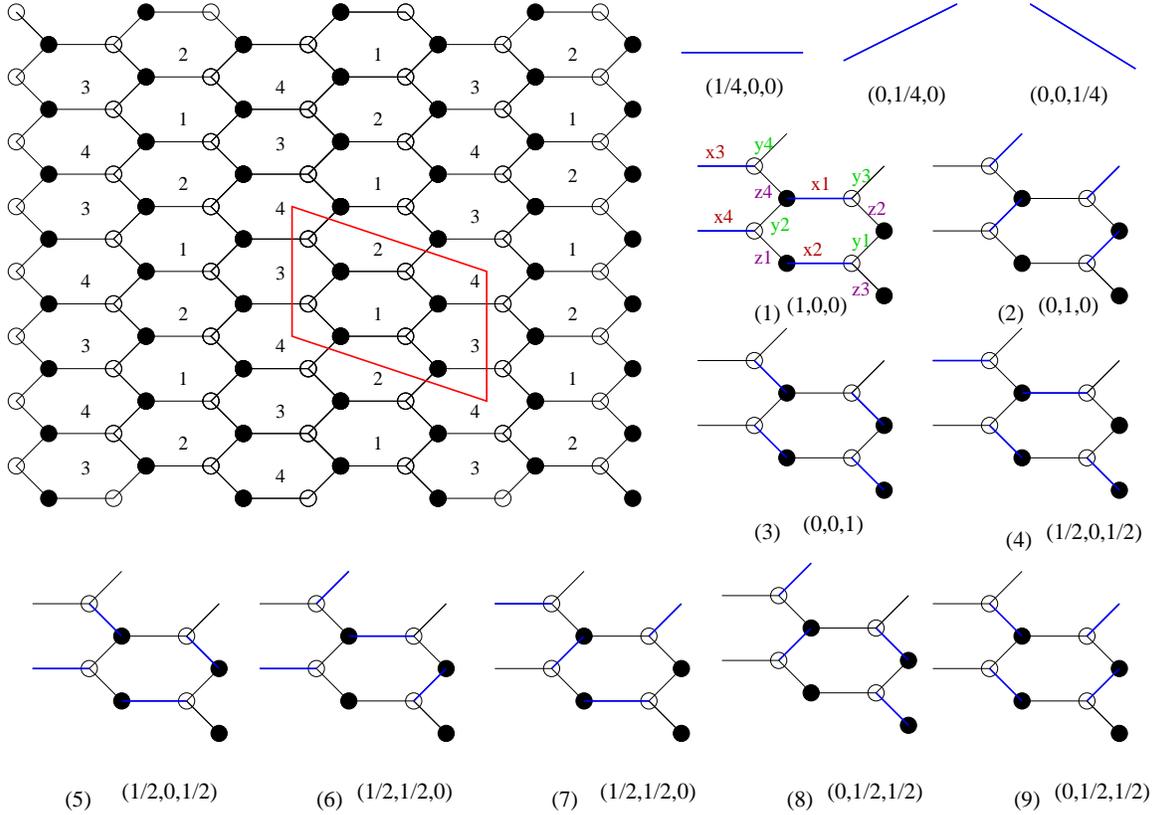}
\caption{\small The fundamental domain and the perfect matchings for the
singularity $\BC^3/\BZ_2\times\BZ_2$. We have also shown the labeling
for the edges, which correspond to bifundamental fields in the gauge
theory.}
\label{fig2}
\end{figure}

Let us begin this subsection with an analysis for the orbifold
$\BC^3/\BZ_2\times\BZ_2$, with the orbifolding action being
\begin{eqnarray}
g_1 : \left(Z_1, Z_2, Z_3\right) \to \left(-Z_1, -Z_2, Z_3\right)  
\nonumber\\
g_1 : \left(Z_1, Z_2, Z_3\right) \to \left(-Z_1, Z_2, -Z_3\right)
\end{eqnarray}
where the $Z_i, i=1,2,3$ denote the coordinates of $\BC^3$.
In the closed string description of this orbifold, we consider the 
$\BZ^{\oplus 3}$ lattice generated by the basis vectors 
${\vec e_1} = (1,0,0)$, ${\vec e_2} = (0,1,0)$ and ${\vec e_3} = (0,0,1)$,
and augment them with the fractional points that correspond to the 
R-charges (of the closed string twist operator) of the three marginal 
sectors of the theory. These are given by the vectors 
${\vec e_4} = (\frac{1}{2},0,\frac{1}{2})$, 
${\vec e_5} = (\frac{1}{2},\frac{1}{2},0)$, 
${\vec e_6} = (0,\frac{1}{2},\frac{1}{2})$ and correspond to the
action by $g_1$, $g_2$ and $g_1.g_2$. The toric diagram for the 
orbifold is shown in fig. (\ref{fig1}). In the same figure, we have also 
shown the position of the lattice vectors ${\vec e_1},\cdots,{\vec e_6}$, 
along with their multiplicities in the brane probe picture \cite{fhh}, 
and a partial resolution of this to the non-orbifold SPP singularity.   
%%%%%%%%%%%%%%%%%%%%%%%%%%%%
%%%%%%%%%%%%%%%%%%%%%%%%%%%%
In fig. (\ref{fig2}), we show the dimer model for this singularity, and
its perfect matchings. The perfect matchings are classified according
to their closed string twisted sector R charges, which can be 
read off by assigning weights to the edges of the original hexagonal 
lattice \cite{tapodimer}. These have also been shown in 
fig. (\ref{fig2}), where we have assigned weights to 
the edges according to their orientation, with the condition that three
different types of edges meet at each vertex. From fig. (\ref{fig2}),
we can read off the matching matrix for the $\BC^3/\BZ_2\times\BZ_2$
singularity, and it is given by 
{\small
\begin{eqnarray}
{\cal M}=
\pmatrix{
~&p_1&p_2&p_3&p_4&p_5&p_6&p_7&p_8&p_9\cr
x1&1&0&0&1&0&1&0&0&0\cr
x2&1&0&0&0&1&0&1&0&0\cr
x3&1&0&0&1&0&0&1&0&0\cr
x4&1&0&0&0&1&1&0&0&0\cr
y1&0&1&0&0&0&1&0&0&1\cr
y2&0&1&0&0&0&0&1&1&0\cr
y3&0&1&0&0&0&0&1&0&1\cr
y4&0&1&0&0&0&1&0&1&0\cr
z1&0&0&1&1&0&0&0&0&1\cr
z2&0&0&1&0&1&0&0&1&0\cr
z3&0&0&1&1&0&0&0&1&0\cr
z4&0&0&1&0&1&0&0&0&1\cr
}\label{c3z2z2}\end{eqnarray}}
where the rows denote the $12$ surviving bifundamental fields in the 
D-brane gauge theory after the orbifold projection, and we have explicitly 
labeled these by the edge numbers appearing in fig. (\ref{fig2}), and
the columns correspond to the perfect matching number which has been
given in that figure.  

%%%%%%%%%%%%%%%%%%%%%%%%%%%%
%%%%%%%%%%%%%%%%%%%%%%%%%%%%
From the closed string perspective,
since the toric diagram in this case does not contain any internal point, 
each face symmetry (three such symmetries are independent as can be seen
from the dimer diagram) must necessarily involve
one corner point of the diagram, and three other points. It is easy to write 
these down, as combinations of the points in the toric diagram that sum 
up to zero closed string R-charge, in the spirit of the last section, 
and one possible choice for these combinations is 
\begin{eqnarray}
{\vec e_1} - {\vec e_4} - {\vec e_5} + {\vec e_6} &=& 0\nonumber\\
{\vec e_2} + {\vec e_4} - {\vec e_5} - {\vec e_6} &=& 0\nonumber\\
{\vec e_3} - {\vec e_4} + {\vec e_5} - {\vec e_6} &=& 0
\label{z2z2ts}
\end{eqnarray}

Now, from the matching matrix ${\cal M}$ of eq. (\ref{c3z2z2}), we can obtain 
its redundancy matrix, which is, in the case, a $3 \times 9$ matrix. 
The set of face symmetries can be obtained directly from ${\cal M}$ 
by noting from fig. (\ref{fig2}) that these symmetries involve the edges
\begin{equation}
F1 : \left(x1,x2,y1,y2,z1,z2\right);~~~ 
F2 : \left(x1,x2,y3,y4,z3,z4\right);~~~
F3 : \left(x3,x4,y1,y2,z3,z4\right)
\end{equation}
Combinations of the perfect matchings that give this
set of edges can be constructed from the columns of the matching matrix
by forming linear combinations whose only nonzero (unit) entries are
at the positions of the above edges. In this case, it can be checked that 
one possible choice of these combinations for the faces $F1, F2, F3$ are
respectively, 
\begin{equation}
p3 - p5 + p6 - p8;~~~p2 + p5 - p6 - p8 ;~~~p1 - p5 - p6 + p8  
\end{equation}  
These are of course equivalent to the combinations in eq. (\ref{z2z2ts}).
The signs are chosen so that each face is traversed in
clockwise sense.  
This information can then be used to construct the quiver diagram for
the singularity $\BC^3/\BZ_2\times\BZ_2$ in a standard manner, following
eq. (\ref{conj1}). 
Let us now study the partial resolutions of the $\BC^3/\BZ_2\times\BZ_2$
singularity. This is interesting, because in this case, 
partial resolutions give rise to massless adjoint fields. 

\begin{figure}[h]
\centering
\epsfxsize=4.5in
\hspace*{0in}\vspace*{.2in}
\epsffile{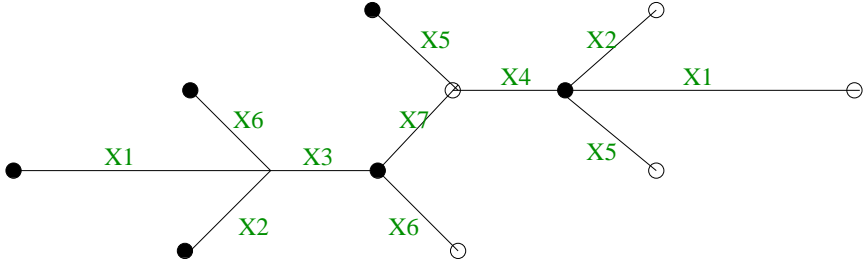}
\caption{\small Dimer model for the SPP singularity.} 
\label{fig3}
\end{figure}

We begin with the matching matrix ${\cal M}$ of $\BC^3/\BZ_2\times\BZ_2$, 
given in eq. (\ref{c3z2z2}). In order to remove one corner of the toric
diagram of fig. (\ref{fig1}), we proceed by removing the row $x1$ of
${\cal M}$. This gives the partial resolution of the parent singularity
to the suspended pinch point (SPP), as shown in fig. (\ref{fig1}). Let us now
understand the combinatorial description of this process. Removing the
edge $x1$ gives us a reduced matching matrix, which can be constructed
by directly deleting the first row and the first, fourth and sixth 
columns of ${\cal M}$. The reduced charge matrix is now given by the
kernel of the reduced matching matrix,
\begin{equation}
Q_{F(r)} = \left(-1,-1,0,0,1,1\right)
\end{equation}
The dual of the kernel of $Q_r$ (which is the $5 \times 6$ matrix $T_r$) 
is given by the set of vectors
{\small
\begin{eqnarray}
{\cal K}_r=
\pmatrix{
0&0&0&0&1&1\cr
0&0&1&1&0&0\cr
0&1&0&0&0&0\cr
1&0&0&0&0&0\cr
0&0&0&1&0&1\cr
}\label{spp1}\end{eqnarray}}
Hence, removing the edge $x1$ has resulted in the removal of $5$ more
edges, i.e a total of $6$ edges out of the initial $12$ have been removed 
(so that we have a resulting graph with six remaining edges). In a
Higgsing procedure, one would expect that a total of five edges
get removed on removing one of the edges of the graph for the parent
singularity. The matrix of eq. (\ref{spp1}) therefore signals the 
appearance of an adjoint field. Note that the superpotential calculated
from ${\cal K}_r$ is inconsistent. Further, if we went ahead with this matrix 
${\cal K}_r$ and calculated the resulting matching matrix, the result would be 
{\small
\begin{eqnarray}
{\cal M}_r=
\pmatrix{
0&0&1&0&0&0\cr
0&0&0&1&0&0\cr
1&0&0&0&1&0\cr
0&1&0&0&1&0\cr
1&0&0&0&0&1\cr
0&1&0&0&0&1\cr
}\label{spp2}\end{eqnarray}}
with now the rows denoting the new edges and the columns labeling the
perfect matchings in the new graph corresponding to the 
SPP singularity. As a perfect matching matrix, eq. (\ref{spp2}) is 
clearly inconsistent. This is because the same number of
bifundamental fields do not appear in each perfect matching. 
The observation here is that we can remedy the situation by inserting 
an extra column in eq. (\ref{spp1}), (corresponding to the adjoint field), 
so that the rows in the matrix ${\cal K}_r$ are forced to add up to 
the same integer ($2$ in this case). On applying this modification, we 
arrive at the matrix
{\small
\begin{eqnarray}
{\cal K}_r'=
\pmatrix{
0&0&0&0&1&1&0\cr
0&0&1&1&0&0&0\cr
0&1&0&0&0&0&1\cr
1&0&0&0&0&0&1\cr
0&0&0&1&0&1&0\cr
}\label{spp3}\end{eqnarray}}
where now the seven columns refer to the seven fields in the theory 
with the last column being the added adjoint. This can be integrated 
to give the superpotential 
\begin{equation}
W_{spp} = X_1X_2X_3X_6 - X_1X_2X_4X_5 + X_3X_6X_7 - X_4X_5X_7
\end{equation}
This is now seen to match with the result of \cite{pru}, with
the $X_i, i=1,\cdots 7$ being the bifundamentals in the D-brane
gauge theory of the SPP singularity, and in terms of these fields, 
the matching matrix for the SPP singularity is seen to be
{\small
\begin{eqnarray}
{\cal M}_{spp}=
\pmatrix{
~&{\tilde p}_1&{\tilde p}_2&{\tilde p}_3&{\tilde p}_4&{\tilde p}_5
&{\tilde p}_6\cr
X_1&0&0&1&0&0&0\cr
X_2&0&0&0&1&0&0\cr
X_3&1&0&0&0&1&0\cr
X_4&0&1&0&0&1&0\cr
X_5&1&0&0&0&0&1\cr
X_6&0&1&0&0&0&1\cr
X_7&0&0&1&1&0&0\cr
}\label{spp4}\end{eqnarray}}
where the rows $Xi, i=1,\cdots,7$ label the seven bifundamentals and 
the columns ${\tilde p}_i, i=1,\cdots,6$ denote the
six perfect matchings of the partially resolved theory. 
From the superpotential above, we may construct the dimer diagram for
the SPP singularity. This is well known, and shown in fig. (\ref{fig3}).
%%%%%%%%%%%%%%%%%%%%%%%%%%%%
%%%%%%%%%%%%%%%%%%%%%%%%%%%%
From the new matching matrix of eq. (\ref{spp4}), we can construct the face
symmetries. From fig. (\ref{fig3}), we see that there are three faces,
so that only two of them give independent constraints. 
The set of edges that participate in these face symmetries
are given by $\left(X_1,X_2,X_3,X_6\right)$ and 
$\left(X_1,X_2,X_4,X_5\right)$. From eq. (\ref{spp4}), when the signs 
are appropriately taken care of (so that the faces are traversed in 
the same sense), these set of edges correspond to the combinations
\begin{equation}
-{\tilde p}_2 + {\tilde p}_3 - 
{\tilde p}_4 + {\tilde p}_5~~~{\rm and}~~~
{\tilde p}_1 - {\tilde p}_3 + {\tilde p}_4 - {\tilde p}_5
\end{equation} 
where the ${\tilde p}_i, i=1,\cdots, 6$ label the columns of ${\cal M}_{spp}$.
Forming the matrix 
{\small
\begin{eqnarray}
A= Q_{D(spp)}=
\pmatrix{
0&-1&1&-1&1&0\cr
1&0&-1&1&-1&0\cr
}\end{eqnarray}}
we obtain the matrix $Q_{D(spp)}.{\cal M}^T_{spp}=d$, from which, removing 
the redundant $U(1)$ gives 
{\small
\begin{eqnarray}
{\Delta}_{spp}=
\pmatrix{
1&-1&1&0&0&-1&0\cr
-1&1&0&-1&1&0&0\cr
}\end{eqnarray}}
which correctly describes the quiver for the SPP. Equivalently, one could
have directly calculated the quiver charges from eq. (\ref{conj1}). 

Let us now move to our next example, where we remove a further edge
from the $\BC^3/\BZ_2\times\BZ_2$ singularity. This can be achieved
conveniently by removing, say, the first row (and thus the third column)
from the matching matrix for the SPP, ${\cal M}_{spp}$, given in 
eq. (\ref{spp4}). It can be checked that this is equivalent to 
directly removing the bifundamentals denoted by $x1$ and $x4$ from the 
matching matrix of $\BC^3/\BZ_2\times\BZ_2$ given in eq. (\ref{c3z2z2}). 
This imples that (from the toric diagram), we are left with the  
singularity $\BC^2/\BZ_2 \times \BC$. The dual of the kernel of the 
redundancy matrix in this case is given by
{\small
\begin{eqnarray}
{\cal K}_{r}''=
\pmatrix{
0&0&0&1&1\cr
0&1&1&0&0\cr
1&0&0&0&0\cr
0&0&1&0&1\cr
}\end{eqnarray}}
The perfect matching matrix here suffers from the same 
inconsistency as before, and hence we introduce one more adjoint 
field to write down the modified matrix
{\small
\begin{eqnarray}
{\cal K}_{\BC^2/\BZ_2\times\BC}= 
\pmatrix{
0&0&0&1&1&0\cr
0&1&1&0&0&0\cr
1&0&0&0&0&1\cr
0&0&1&0&1&0\cr
}\end{eqnarray}}
which gives us the superpotential
\begin{equation}
W_{\BC^2/\BZ_2\times\BC}= X_1X_2X_5 - X_1X_3X_4 + X_2X_5X_6 - X_3X_4X_6
\end{equation}
A computation analogous to that for the SPP (or using eq. 
(\ref{conj1})) now yields the quiver charge matrix
\begin{equation}
\Delta_{\BC^2/\BZ_2\times\BC}=\left(0,1,-1,1,-1,0\right)
\end{equation}
which tells us that the fields $X_1$ and $X_6$ are adjoints. 

Before we conclude this subsection, we briefly comment on the resolution
of the $\BC^3/\BZ_2\times\BZ_2$ singularity to the conifold. As is known,
this theory does not have F-terms and the entire information of the gauge
theory is contained in the D-term equations. We can see this from the
matching matrix of eq. (\ref{c3z2z2}). In order to reach the conifold
singularity from the original matching matrix, we can remove the 
bifundamentals denoted by $x1$ and $y1$. Eq. (\ref{c3z2z2}) then
tells us that along with these fields, five of the nine perfect matchings
have to be removed, and we are left with a theory that has four perfect
matchings. However, the redundancy matrix corresponding to the 
$7 \times 4$ matrix that remain after this removal is the null matrix
$0_{1\times 4}$. Hence, we can take the $T_r$ matrix in this case as the 
matrix ${\rm Id}_{4\times 4}$, whose dual matrix ${\cal K}_r$ is again the
$4 \times 4$ identity matrix. The matching matrix for the conifold
is thus ${\rm Id}_{4\times 4}$. The dimer covering and the perfect matchings
can now be constructed in a standard way, and eq. (\ref{conj1})
can be implemented in a standard manner to give rise to the charge matrix
{\small
\begin{eqnarray}
\Delta_{conifold}= 
\pmatrix{
1&-1&-1&1\cr
}\end{eqnarray}}
As a final remark, note that at each stage, the toric diagrams of the 
resolutions can be obtained by constructing the perfect matchings, whose
information is contained in the matching matrix, and constructing their
height functions. Equivalently, this can be done by directly removing
points from that of the parent singularity corresponding to the
bifundamentals that are removed. 
     
We now discuss the singularity $\BC^2/\BZ_2\times\BZ_3$. This is the next
nontrivial example where adjoint fields appear on partial resolutions 
of the singularity. 

\subsection{The Singularity $\BC^3/\BZ_2\times\BZ_3$}

\begin{figure}[h]
\centering
\epsfxsize=2.0in
\hspace*{0in}\vspace*{.2in}
\epsffile{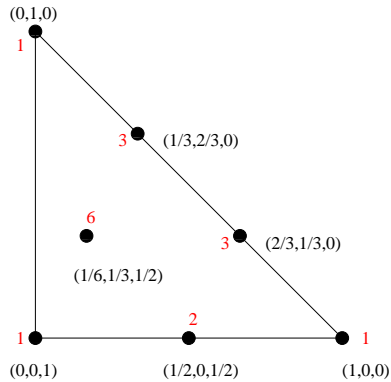}
\caption{\small Toric diagram for the singularity $\BC^3/\BZ_2\times\BZ_3$.
The closed string R-charges are shown, along with the multiplicities of
the fields in the open string picture.} 
\label{fig4}
\end{figure}

\begin{figure}[h]
\centering
\epsfxsize=3.5in
\hspace*{0in}\vspace*{.2in}
\epsffile{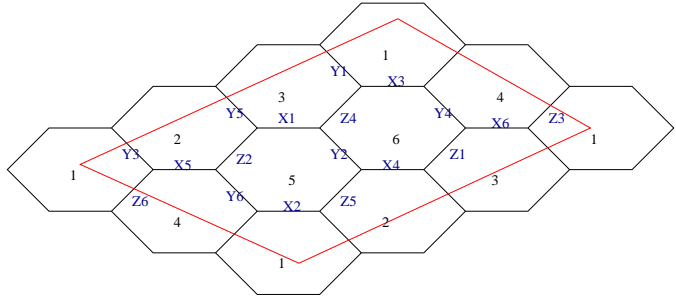}
\caption{\small Fundamental region for the dimer model of the singularity
$\BC^3/\BZ_2\times\BZ_3$. We have labeled the edges in accordance with
the previous subsection.} 
\label{fig5}
\end{figure}

\begin{figure}[h]
\centering
\epsfxsize=5.0in
\hspace*{0in}\vspace*{.2in}
\epsffile{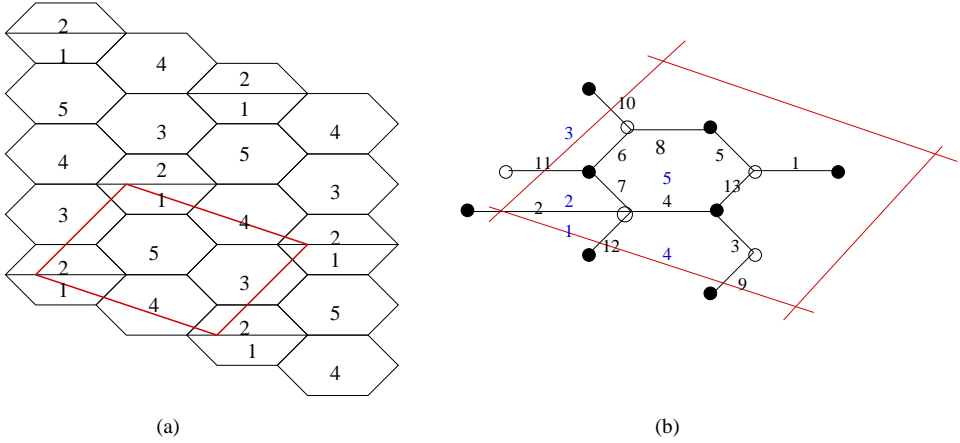}
\caption{Dimer covering for the gauge theory for the orbifold 
$\BC^3/\BZ_2\times\BZ_3$ with the edge $X_1$ removed. In (a), we
have shown the fundamental domain of this covering. (b) shows a 
labeling of the bifundamental fields corresponding to the superpotential
in eq. (\ref{sup1}) (in black). The blue colored integers in (b) refer
to the original labeling of the faces in (a).} 
\label{fig6}
\end{figure}

In this subsection, we study the orbifold $\BC^2/\BZ_2\times\BZ_3$, where
the orbifolding group implies an asymmetric action on the coordinates.
We will see how the computational tools introduced in the last subsection
can be effectively used in this context as well. The closed
string description of this orbifold parallels the one discussed
in the last subsection. Specifically, we choose the action of the orbifolding
group on the coordinates as 
\begin{eqnarray}
g_1 = \left(Z_1,Z_2,Z_3\right) \to \left(-Z_1,Z_2,-Z_3\right)\nonumber\\
g_2 = \left(Z_1,Z_2,Z_3\right) \to \left(\omega Z_1,\omega^2 Z_2,Z_3\right)
\end{eqnarray}
where $\omega = e^{\frac{2\pi i}{3}}$ and the $Z_i, i=1,2,3$ are the
coordinates of $\BC^3$. Taking into account the 
various marginal twisted sectors (in addition to the generators of an
$SL(3,Z)$ lattice), we obtain the 
toric diagram shown in fig. (\ref{fig4}) (after projection to a convenient
plane, so that the vectors are coplanar). In fig. (\ref{fig4}), 
we have also marked the closed string R-charges.  
%%%%%%%%%%%%%%%%%%%%%%%%%%%%
%%%%%%%%%%%%%%%%%%%%%%%%%%%%
The fundamental region for the dimer model for this singularity is shown 
in fig. (\ref{fig5}). 
%%%%%%%%%%%%%%%%%%%%%%%%%%%%
%%%%%%%%%%%%%%%%%%%%%%%%%%%%
In the appendix, we have shown diagrammatically the $17$ possible perfect 
matchings for the dimer model of this singularity. These can be obtained
by using the Kasteleyn matrix, as in the last section. Also, in the 
appendix, we have provided
the matching matrix for this singularity in eq. (\ref{c3z2z31}), where
the $18$ surviving bifundamental fields of this theory have been collectively
labelled as $X_i, i=1,\cdots,18$.
From this matrix, we can see that due to the asymmetric action of the
orbifold, removing different nodes may result in the removal of 
different number of perfect matchings during Higgsing.  
Again, the face symmetries correspond to the total closed string R-charge
vanishing around each face. These are also seen to be combinations only
of the perfect matchings corresponding to the internal point in the
toric diagram of fig. (\ref{fig4}).
Let us now consider some blowups of this singularity, via the Higgsing
procedure from the dimer model perspective. These have been previously
considered in \cite{pru}. First, we give a vev to one of the fields, 
in eq. (\ref{c3z2z31}), say $X_1$. The charge matrix for the reduced
singularity is found to be
{\small
\begin{eqnarray}
Q^1_{F(r)}= 
\pmatrix{
-1&1&0&0&0&-1&-1&1&0&0&1\cr
0&-1&0&0&-1&0&1&0&0&1&0\cr
1&-1&-1&0&-1&1&1&-1&1&0&0\cr
1&-1&-1&1&0&0&0&0&0&0&0\cr
}\label{c3z2z3rmx12}\end{eqnarray}}
This gives rise to the reduced $K$ matrix, now with $13$ fields,
{\small
\begin{eqnarray}
{\cal K}^1_r=
\pmatrix{
0&0&0&0&0&0&0&0&0&1&1&1&1\cr 
0&0&0&0&0&0&1&1&1&0&0&0&1\cr 
0&0&0&1&1&1&0&0&1&0&0&0&0\cr
0&1&1&0&1&1&0&0&0&0&0&0&0\cr 
1&0&1&0&0&1&0&0&0&0&0&1&0\cr 
0&1&0&0&0&0&0&1&0&0&1&0&1\cr 
0&0&0&0&0&1&0&0&1&0&0&1&1\cr 
}\label{c3z2z3rmx12k}  
\end{eqnarray}}
which can be integrated to give the superpotential
\begin{eqnarray}
W = &~& Y_1Y_5Y_{13} - Y_3Y_4Y_{13} + Y_3Y_9Y_{11} - Y_5Y_8Y_{12}
- Y_6Y_7Y_{11} + Y_6Y_8Y_{10}\nonumber\\
&-& Y_1Y_2Y_9Y_{10} + Y_2Y_4Y_7Y_{12}
\label{sup1}
\end{eqnarray}
where we have labeled the new fields as $Y_i, i=1,\cdots 13$ to avoid
confusion. The reduced matching matrix now involves
$13$ bifundamentals and $11$ perfect matchings, and is given by 
\begin{eqnarray}
{\cal M}^1_r=
\pmatrix{
~&{\tilde p}_1&{\tilde p}_2&{\tilde p}_3&{\tilde p}_4&{\tilde p}_5
&{\tilde p}_6&{\tilde p}_7&{\tilde p}_8&{\tilde p}_9
&{\tilde p}_{10}&{\tilde p}_{11}\cr
Y_1&0&1&0&1&0&0&1&0&0&0&0\cr Y_2&0&0&0&0&0&1&0&1&0&0&0\cr 
Y_3&1&1&0&0&0&0&1&1&0&0&0\cr Y_4&0&0&1&1&0&0&0&0&1&0&0\cr 
Y_5&1&0&1&0&0&0&0&1&1&0&0\cr Y_6&0&0&0&0&1&0&1&1&1&0&0\cr
Y_7&1&1&0&0&0&0&0&0&0&1&0\cr Y_8&0&1&0&1&0&1&0&0&0&1&0\cr
Y_9&0&0&0&0&1&0&0&0&1&1&0\cr Y_{10}&1&0&1&0&0&0&0&0&0&0&1\cr
Y_{11}&0&0&1&1&0&1&0&0&0&0&1\cr Y_{12}&0&0&0&0&1&0&1&0&0&0&1\cr
Y_{13}&0&0&0&0&1&1&0&0&0&1&1\cr
}\label{c3z2z3x12rem}\end{eqnarray}
Where now the ${\tilde p}_i, i = 1\cdots 11$ are the new perfect matchings
that descend from the parent theory. 
The dimer model for this partial resolution of $\BC^3/\BZ_2\times\BZ_3$
is shown in fig. \ref{fig6} a. 
%%%%%%%%%%%%%%%%%%%%%%%%%%%%
%%%%%%%%%%%%%%%%%%%%%%%%%%%%
From fig. (\ref{fig6} a), we can directly draw the toric diagram, or, in
the spirit of the previous subsection, note that the five distinct faces
of the dimer diagram in the figure are generated by the bifundamentals
\begin{eqnarray}
F1 &=& \left(Y_2,Y_8,Y_{10},Y_{12}\right),~~
F2 = \left(Y_2,Y_7,Y_9,Y_{11}\right),~~
F3 = \left(Y_1,Y_3,Y_6,Y_{10},Y_{11},Y_{13}\right),\nonumber\\
F4 &=& \left(Y_1,Y_3,Y_4,Y_5,Y_9,Y_{12}\right),~~
F5 = \left(Y_4,Y_5,Y_6,Y_7,Y_8,Y_{13}\right)
\end{eqnarray}
This implies that the face symmetries are generated by the matrix
\begin{eqnarray}
Q_D^1=
\pmatrix{
~&1&0&0&0&0&-1\cr
~&-1&0&0&0&1&0\cr
0_{5\times5}&0&-1&0&0&0&1\cr
~&0&1&0&-1&0&0\cr
~&0&0&0&1&-1&0\cr
}
\label{c3z2z3q1}
\end{eqnarray}
from which we read off the quiver matrix
\begin{eqnarray}
d=
\pmatrix{
1&0&1&-1&-1&0&0&0&-1&0&0&1&0\cr
-1&0&-1&0&0&-1&0&0&0&1&1&0&1\cr
0&0&0&1&1&1&-1&-1&0&0&0&0&-1\cr 
0&1&0&0&0&0&0&1&0&-1&0&-1&0\cr
0&-1&0&0&0&0&1&0&1&0&-1&0&0\cr
}\end{eqnarray}

\begin{figure}[h]
\centering
\epsfxsize=1.7in
\hspace*{0in}\vspace*{.2in}
\epsffile{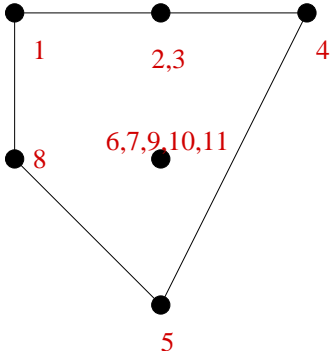}
\caption{Toric diagram for the orbifold $\BC^3/\BZ_2\times\BZ_3$ with 
the edge $X_1$ removed. The integers in red label the perfect matchings
that appear in eq. (\ref{c3z2z3x12rem}). The positions of these 
correspond to the height functions as shown in fig. (\ref{new2b}) in 
the appendix.} 
\label{new3}
\end{figure}

Now we need to check that the face symmetries are indeed obtainable from
purely internal matchings of the toric diagram for this dimer model. 
This is not difficult to see. In fig. (\ref{fig6} b),
we have shown (one choice for) the labeling of the fields that arise
in the superpotential of eq. (\ref{sup1}). In fig. (\ref{new2a}) and 
(\ref{new2b}) in the appendix, we have
shown explicitly the eleven perfect matchings corresponding to the matching
matrix in eq. (\ref{c3z2z3x12rem}). From the labeling of the matchings
in terms of the height functions, we see that ${\tilde p}_6$, ${\tilde p}_7$,
${\tilde p}_9$, ${\tilde p}_{10}$ and ${\tilde p}_{11}$ are the five internal
matchings, as expected from eq. (\ref{c3z2z3q1}). The toric diagram for 
this partial resolution is shown in fig. (\ref{new3}).  
%%%%%%%%%%%%%%%%%%%%%%%%%%%%
%%%%%%%%%%%%%%%%%%%%%%%%%%%%

Consider now a further blowup of this orbifold, wherein we give a vev
to the field $Y_1$ in eq. (\ref{c3z2z3x12rem}). The reduced charge matrix
is given by 
\begin{eqnarray}
Q^2_{F(r)}=\pmatrix{
0&-1&-1&0&0&1&0&1\cr
-1&1&0&-1&1&-1&1&0\cr
}\end{eqnarray}
from which we can directly see that this is the cone of the first del Pezzo
surface $dP_1$ \cite{masterspace}, equivalently, this conclusion can be
reached by calculating the superpotential which, in this case, can 
be determined as  
\begin{equation}
W = Y_1Y_7Y_{10} - Y_2Y_7Y_8 + Y_4Y_6Y_8 - Y_4Y_5Y_{10} + 
Y_2Y_3Y_5Y_9 - Y_1Y_3Y_6Y_9
\end{equation} 
with the $Y_i, i=1,\cdots 10$ now denoting the bifundamental fields 
of the new theory. 
Now, consider giving a vev to the field $Y_2$ in eq. (\ref{c3z2z3x12rem}).
In this case, we can calculate the reduced charge matrix to be
\begin{eqnarray}
Q^3_{F(r)}=\pmatrix{
0&0&-1&0&-1&0&1&0&1\cr
0&-1&0&0&-1&1&0&1&0\cr
1&-1&-1&1&0&0&0&0&0\cr
}\label{c3z4}\end{eqnarray}
This singularity has a toric diagram which can be obtained from that 
in fig. (\ref{new3}) with the points labeled $6$ and $8$ removed, i.e it
is a triangle with one internal point (with multiplicity $4$) and one point 
on the boundary (with multiplicity $2$). It is known that such toric diagrams with
points on the boundary typically arise in orbifolds which might have a non-isolated
singularity \cite{muto}. Further, in this case, from the charge matrix of 
eq. (\ref{c3z4}), the partially resolved theory is seen to contain $12$
bifundamental fields. By explicit computation using the forward algorithm, we 
have checked that this is in fact the  $\BC^3/\BZ_4$ singularity. The 
superpotential obtained from the ${\cal K}_r$ matrix computed from
eq. (\ref{c3z4}) also confirms this.   

Before we end this subsection, let us consider one more example. 
Consider removing the edge $Z_4$ from
the matching matrix of $\BC^3/\BZ_2\times\BZ_3$. It can be checked that
this results in an inconsistent superpotential for the resulting 
theory. Now, we consider removing the bifundamentals $Z_4$ and 
$Z_6$ simultaneously. This yields the reduced ${\cal K}_r$ matrix, in which
we need to add two adjoint fields (to make the resulting matching 
matrix consistent), and after this, the matrix reads
\begin{eqnarray}
{\cal K}^4_r = \pmatrix{
0&0&0&0&0&0&0&1&1&1\cr
0&0&0&0&0&0&1&0&1&1\cr
0&0&0&1&1&1&0&0&0&0\cr
0&1&1&0&0&1&0&0&0&0\cr
1&0&1&0&1&0&0&0&0&0\cr
0&1&0&1&0&1&0&0&0&0\cr
}\label{puzzle}\end{eqnarray}
where the last two columns denote the (adjoint) fields that we have added.
For this case, we obtain the superpotential
\begin{equation}
W = Y_1Y_6Y_9 - Y_2Y_5Y_9 + Y_3Y_4Y_{10} - Y_1Y_6Y_{10} 
+ Y_2Y_5Y_7Y_8 - X_3Y_4Y_7Y_8
\end{equation}
where $Y_i, i=1,\cdots 10$ denote the bifundamental fields of the
resulting theory. This is seen to match with the corresponding 
result of \cite{pru} after an obvious
identification of fields. The quiver charges and 
the dimer model of this theory can be obtained by standard means outlined
before, and we do not present them here. Rather, let us point out 
a puzzle that we are unable to resolve at this stage. 
\footnote{This example has appeared in \cite{tapo3}, although at that
stage, it was not known how to rule out inconsistent toric diagrams.}
Suppose we give a vev 
to the fields $Z_1$ and $Z_2$ instead. It can be checked that this 
gives rise to exactly the same ${\cal K}_r$ matrix as in eq. (\ref{puzzle}), and 
hence to the same superpotential with two adjoints. But now the
toric diagram of the resulting theory has a multiplicity at an 
external point, and hence is
inconsistent. In fact, from an algebraic analysis, this singularity
can be shown to have an equation that is analogous to the  
singularity $\BC^2/\BZ_3\times\BC$, which, in the $N=1$ description, should
have a matter content of $9$ fields, including $3$ adjoints, which is
clearly not the case at hand. The forward algorithm is less useful here, 
due to the presence of adjoint fields, and apriori we do not know how to 
rule out this case. We believe that this might be a generic feature of 
orbifolds of the form $\BC^3/\BZ_m\times\BZ_n$ with $m\neq n$, where removing 
different bifundamentals may remove different numbers of perfect 
matchings, and it needs to be investigated further.

\subsection{The Orbifold $\BC^3/\BZ_3\times\BZ_3$}

\begin{figure}[h]
\centering
\epsfxsize=3.0in
\hspace*{0in}\vspace*{.2in}
\epsffile{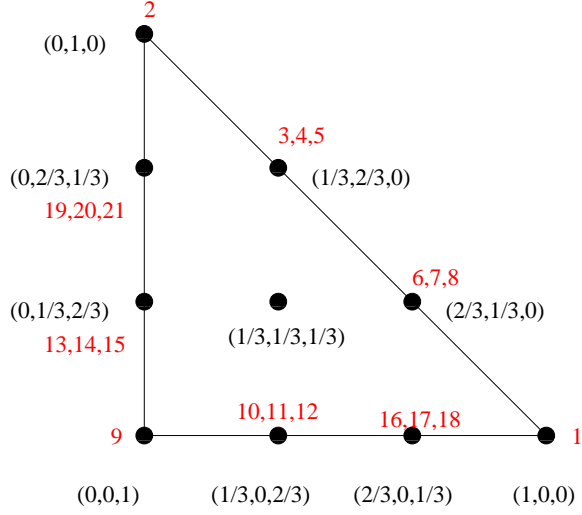}
\caption{\small Toric diagram for the resolution of the 
singularity $\BC^3/\BZ_3\times\BZ_3$. We have shown the closed string 
R-charges of the of the twisted sectors, as well as the perfect 
matchings which correspond to them, following our diagrams for this 
singularity presented in the appendix.}
\label{c3z3z3toric}
\end{figure}

\begin{figure}[h]
\centering
\epsfxsize=3.0in
\hspace*{0in}\vspace*{.2in}
\epsffile{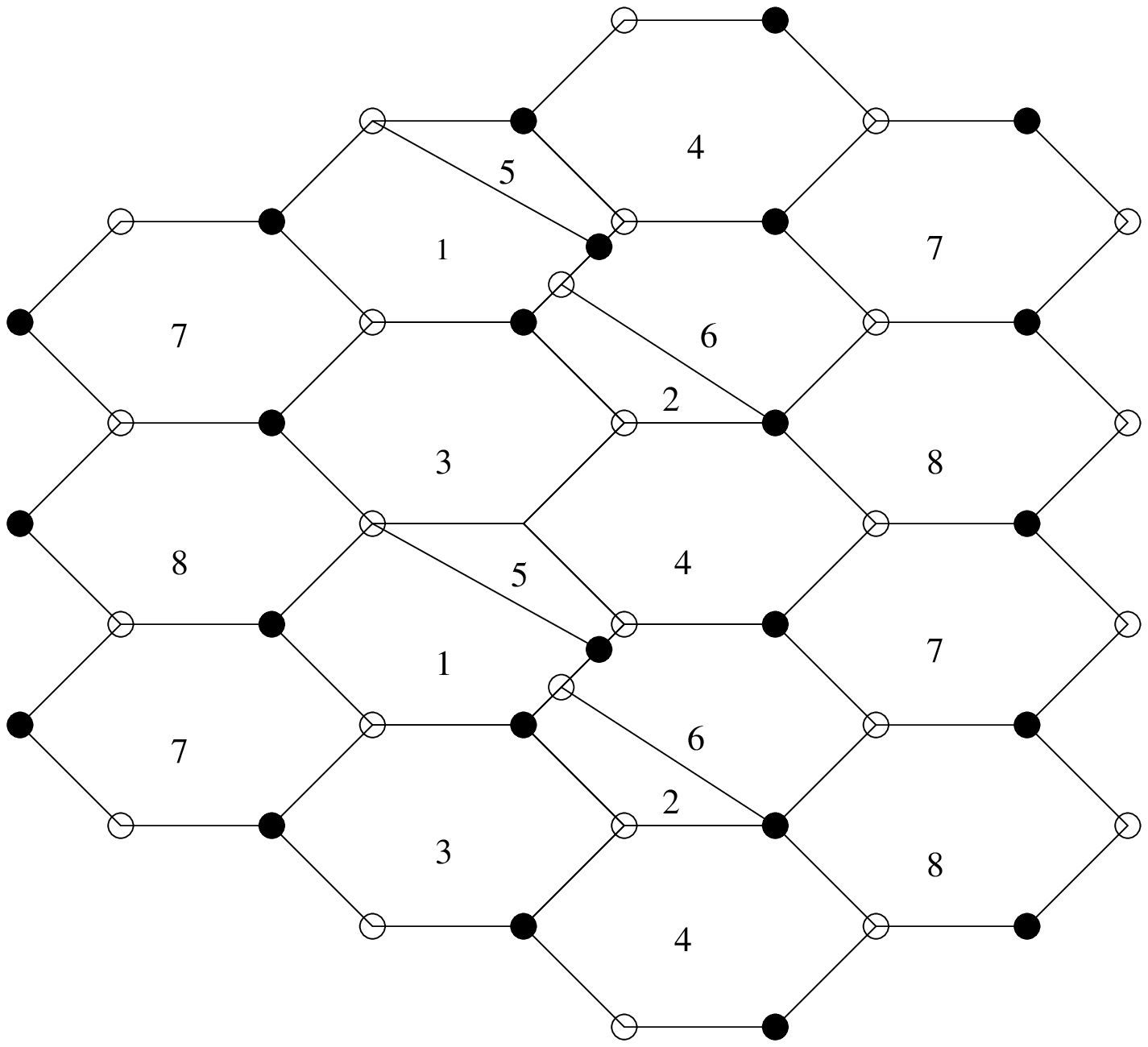}
\caption{\small Dimer model of the orbifold $\BC^3/\BZ_3\times\BZ_3$ with
one bifundamental field removed.} 
\label{c3z3z3two}
\end{figure}

We now present our results on the orbifold $\BC^3/\BZ_3\times\BZ_3$. We
will be brief here due to space constraints, and also becos 
our methods of the previous subsection carry
over straightforwardly in this case (we do not have any complications due
to adjoint fields here). The orbifolding action is 
\begin{eqnarray}
g_1 = \left(Z_1,Z_2,Z_3\right) \to 
\left(\omega Z_1, \omega^2 Z_2, Z_3\right)\nonumber\\
g_2 = \left(Z_1,Z_2,Z_3\right) \to
\left(\omega Z_1, Z_2, \omega^2 Z_3\right)
\end{eqnarray}
where $\omega=e^{\frac{2\pi i}{3}}$ and as usual, $Z_i, i=1,2,3$ denote 
the coordinates of $\BC^3$.
The orbifold projection can be carried out in a standard way, and 
shows that in this case, there are $27$ fields surviving the 
orbifolding action, and that there are 
$42$ perfect matchings. Due to space constraints, we have presented only 
a subset of these matchings in the appendix. In the appendix, we have
also presented the matching matrix for this orbifold. 
From this, one can construct the matching matrix
for the completely singular variety, or its partial resolutions. We have
again checked that the face symmetries are generated by the
perfect matchings that correspond to only internal point in the
toric diagram for this singularity, shown in fig. (\ref{c3z3z3toric}).
In the closed string description, we consider the $\BZ^{\oplus 3}$ lattice, 
generated by the vectors 
${\vec e_1} = \left(1,0,0\right)$, ${\vec e_2} = \left(0,1,0\right)$,
${\vec e_3} = \left(0,0,1\right)$, and include the following
seven fractional points ${\vec e_4}, \cdots,{\vec e_{10}}$ which correspond
to the seven marginal sectors in the theory :
\begin{eqnarray}
&~&\left(\frac{1}{3},\frac{2}{3},0\right),
~\left(\frac{2}{3},\frac{1}{3},0\right),
~\left(\frac{1}{3},0,\frac{2}{3}\right),
~\left(\frac{2}{3},0,\frac{1}{3}\right),\nonumber\\
&~&\left(0,\frac{1}{3},\frac{2}{3}\right),
~\left(0,\frac{2}{3},\frac{1}{3}\right),
~\left(\frac{1}{3},\frac{1}{3},\frac{1}{3}\right)
\end{eqnarray}
In fig. (\ref{c3z3z3toric}), we have also shown the perfect matchings 
corresponding to each closed string twisted sector.  
%%%%%%%%%%%%%%%%%%%%%%%%%%%%
%%%%%%%%%%%%%%%%%%%%%%%%%%%%
We now discuss some resolutions of this singularity. From the  
matching matrix for this orbifold presented in the appendix, it can be 
seen that removal any one edge from the graph removes seven internal and
seven external points from the toric diagram. Removing, for example, 
the bifundamental field $X_1$ can be seen to give rise to a theory
that has a $18\times 24$ charge matrix, from which the (reduced) $T$ matrix  
can be calculated,   
%\begin{eqnarray}
%T^r = \pmatrix{
%-1&-1&0&0&0&0&1&-1&0&0&1&-1&-1&0&-1&0&0&0&0&0&0&0&0&0&0&0&0&1\cr
%2&1&1&0&-2&-1&-1&0&-2&0&0&0&2&0&0&1&0&0&-2&0&-1&-1&0&-1&0&0&1&0\cr
%0&1&0&1&0&0&0&0&0&0&0&0&0&0&1&0&0&0&1&0&0&1&0&1&0&1&0&0\cr
%-1&-1&-1&-1&2&1&1&1&2&0&0&1&-1&0&0&-1&0&0&1&0&1&0&0&1&1&0&0&0\cr 
%-1&-1&0&0&1&1&0&0&1&0&0&0&-1&0&0&-1&0&0&1&0&1&1&1&0&0&0&0&0\cr 
%0&1&0&1&-1&0&0&-1&-1&0&1&-1&0&0&0&1&0&0&0&1&0&0&0&0&0&0&0&0\cr
%0&0&0&0&1&1&0&1&1&0&0&1&0&0&1&0&0&1&0&0&0&0&0&0&0&0&0&0\cr 
%1&1&1&1&-1&-1&0&0&-1&0&0&0&1&0&0&1&1&0&0&0&0&0&0&0&0&0&0&0\cr 
%2&1&1&0&-1&-1&-1&0&0&0&-1&1&1&1&0&0&0&0&0&0&0&0&0&0&0&0&0&0\cr
%-1&-1&-1&-1&2&1&1&1&1&1&0&0&0&0&0&0&0&0&0&0&0&0&0&0&0&0&0&0\cr
%}\end{eqnarray}
and the dual to this matrix is the reduced ${\cal K}_r$ matrix, given by
\begin{eqnarray}
{\cal K}_r=\pmatrix{
0& 0& 0& 0& 0& 0& 0& 0& 0& 0& 0& 0& 0& 0& 0& 1& 1& 1& 1& 1& 1& 1\cr 
0& 0& 0& 0& 0& 0& 0& 0& 0& 0& 0& 1& 1& 1& 1& 0& 0& 0& 0& 1& 1& 1\cr 
0& 0& 0& 0& 0& 0& 0& 0& 0& 0& 1& 0& 0& 1& 1& 0& 0& 0& 1& 1& 1& 1\cr 
0& 0& 0& 0& 0& 0& 0& 1& 1& 1& 0& 1& 1& 0& 1& 0& 0& 0& 0& 0& 0& 1\cr
0& 0& 0& 0& 0& 1& 1& 0& 0& 0& 0& 1& 1& 1& 0& 0& 0& 0& 0& 1& 1& 0\cr 
0& 0& 0& 1& 1& 0& 1& 0& 1& 1& 0& 0& 1& 0& 0& 0& 0& 1& 0& 0& 0& 0\cr 
0& 1& 1& 0& 1& 0& 0& 0& 0& 0& 0& 0& 0& 0& 0& 1& 1& 1& 0& 0& 1& 0\cr 
1& 0& 1& 0& 0& 1& 0& 1& 0& 1& 0& 1& 0& 0& 0& 0& 1& 0& 0& 0& 0& 0\cr 
0& 0& 0& 1& 0& 0& 1& 0& 1& 0& 0& 0& 0& 0& 0& 1& 0& 1& 1& 1& 0& 0\cr 
1& 0& 0& 1& 0& 0& 1& 0& 0& 0& 0& 0& 0& 1& 0& 0& 0& 1& 1& 1& 0& 0\cr
}\end{eqnarray}
and is seen to give the superpotential 
\begin{eqnarray}
W &=& X_1X_9X_{21} - X_2X_{10}X_{20} - X_3X_7X_{22} + X_3X_{13}X_{19}
- X_4X_8X_{21} + X_5X_8X_{20} - X_5X_{12}X_{19} \nonumber \\
&-& X_6X_{15}X_{18} + X_7X_{15}X_{17} - X_9X_{14}X_{17} + X_{10}X_{14}X_{16} 
+ X_{11}X_{12}X_{18} \nonumber\\
&-& X_1X_{11}X_{13}X_{16} + X_2X_4X_6X_{22}
\end{eqnarray}
This gives the dimer covering, which is shown in fig. (\ref{c3z3z3two}).
%%%%%%%%%%%%%%%%%%%%%%%%%%%%
%%%%%%%%%%%%%%%%%%%%%%%%%%%%
Removal of the edge $X_1$ removes one corner of the toric diagram, 
as above, and in order to remove a further corner, from the matching matrix 
for the orbifold provided in the appendix, it can be seen that there are 
fourteen choices for the same. We will just provide the result here. 
We find that $12$ out of these $14$ choices give rise
to consistent superpotentials. These correspond to two of the phases of
the $PdP_4$ theory (which is the third del Pezzo surface blown up at a 
generic point) \cite{unhigdp}. Specifically, for $8$ cases, we get a toric
diagram with $9$ internal points and for $4$ cases, we get a toric diagram
with $12$ internal points, the number of external points being $7$ for
both. The remaining $2$ choices are seen to give inconsistent 
superpotentials, although one cannot apriori rule out
these cases simply from their toric diagrams. The third phases of 
$PdP_4$ cannot be obtained from this procedure, as pointed out in
\cite{unhigdp}. Now, one can remove a further corner from the resulting
toric diagram, and this straightforwardly gives rise to the four phases 
of the $dP_3$ theory. At each stage, starting from the matching matrix for the 
$\BC^3/\BZ_3\times\BZ_3$ orbifold, the charges for the masterspaces of
these theories and their superpotentials can be read off. 

\begin{figure}[h]
\centering
\epsfxsize=5.5in
\hspace*{0in}\vspace*{.2in}
\epsffile{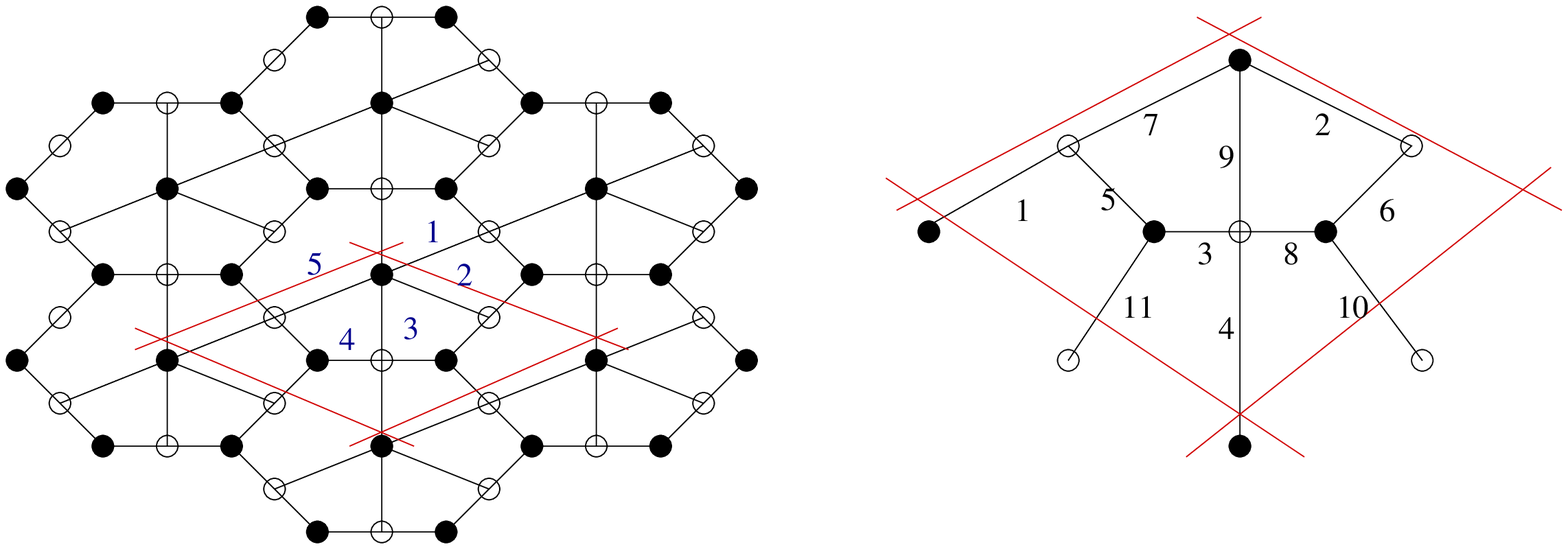}
\caption{The dimer model and fundamental domain for the second del Pezzo
surface is shown on the left. The blue integers label the faces. 
The fundamental domain and the labeling of the fields according
to the superpotential in eq.(\ref{supdp2}) is shown on the right.}
\label{dp2}
\end{figure}

Before ending this section, we present a final example that will also 
substantiate our second conjecture of section (2). We consider the blowup
of the $\BC^3/\BZ_3\times\BZ_3$ orbifold to model II of the second 
del Pezzo surface $dP_2$, which has been well studied in the literature, 
from other perspectives \cite{francovegh}.
From the matching matrix for the orbifold $\BC^3/\BZ_3\times\BZ_3$
presented in the appendix, we remove the bifundamental fields $X_1$,
$X_2$, $Z_2$ and $Z_3$. This gives us a theory with five internal
and five external points. The reduced matrix $T_r$ has, in this case, 
dimensions $7 \times 10$ and its dual is the matrix
{\small 
\begin{eqnarray}
{\cal K}_r=\begin{pmatrix}
{
0&0&0&0&0&0&0&0&1&1&1\cr
0&0&0&0&0&0&1&1&0&0&1\cr
0&0&0&1&1&1&0&0&0&0&0\cr
0&1&1&0&0&0&0&0&0&1&0\cr
1&0&1&0&0&1&0&0&0&0&0\cr
1&0&0&0&0&0&0&1&0&0&1\cr
0&0&0&0&1&1&0&0&1&0&0}
\end{pmatrix}
\end{eqnarray}}
We thus obtain a theory with eleven bifundamentals, which we recognise to
be model II of $dP_2$, and the matrix 
${\cal K}_r$ can be integrated to give the superpotential of the theory
\begin{equation}
W_{dP_2} = X_2X_6X_{11} - X_3X_5X_{11} - X_6X_8X_{10} + X_1X_5X_7X_{10} 
+ X_3X_4X_8X_9 - X_1X_2X_4X_7X_9 
\label{supdp2}
\end{equation}
The reduced matching matrix can be calculated to be
\begin{eqnarray}
{\cal M}_r=\begin{pmatrix}
{
~&{\tilde p}_1&{\tilde p}_2&{\tilde p}_3&{\tilde p}_4&{\tilde p}_5
&{\tilde p}_6&{\tilde p}_7&{\tilde p}_8&{\tilde p}_9&{\tilde p}_{10}\cr
X_1&0&0&0&0&1&1&0&0&0&0\cr
X_2&0&0&1&0&0&0&1&0&0&0\cr
X_4&0&1&0&0&0&1&1&0&0&0\cr
X_4&1&0&0&0&0&0&0&1&0&0\cr
X_5&0&0&1&1&0&0&0&1&0&0\cr
X_6&0&1&0&1&0&1&0&1&0&0\cr
X_7&0&1&0&0&0&0&0&0&1&0\cr
X_8&0&0&1&0&1&0&0&0&1&0\cr
X_9&0&0&0&1&0&0&0&0&0&1\cr
X_{10}&1&0&0&0&0&0&1&0&0&1\cr
X_{11}&1&0&0&0&1&0&0&0&1&1\cr
}
\label{pmdp2}
\end{pmatrix}
\end{eqnarray}
where ${\tilde p}_i, i = 1,\cdots,10$ label the ten perfect matchings of
the $dP_2$ singularity.  
From the superpotential in eq. (\ref{supdp2}), the dimer covering 
corresponding to this model can be obtained. This is well know, and we 
reproduce it in fig. (\ref{dp2}), where we have
also shown the fundamental region and the labeling of the fields
obtained from eq. (\ref{supdp2}).  
%%%%%%%%%%%%%%%%%%%%%%%%%%%%
%%%%%%%%%%%%%%%%%%%%%%%%%%%%
From fig. (\ref{dp2}), we see that the faces in the dimer diagram are
generated by the bifundamental fields 
\begin{eqnarray}
&~& F1 = \left(X_1,X_4,X_8,X_{10}\right),~ F2 = \left(X_1,X_2,X_5,X_{11}\right),
~F3 = \left(X_2,X_6,X_8,X_9\right),\nonumber\\
&~& F4 = \left(X_3,X_5,X_7,X_9\right),~F5 = \left(X_3,X_4,X_6,X_7,X_{10},X_{11}
\right)
\end{eqnarray}
and these are, from eq. (\ref{pmdp2}), given by the perfect matchings
${\tilde p}_1 - {\tilde p}_5$, ${\tilde p}_5 - {\tilde p}_3$, 
${\tilde p}_3 - {\tilde p}_4$, ${\tilde p}_4 - {\tilde p}_2$, 
${\tilde p}_2 - {\tilde p}_1$
Hence, we need to check that the perfect matchings ${\tilde p}_i,
i=1,\cdots 5$ are now the internal matchings. In fig. (\ref{dp2pm})
in the appendix, we have provided the ten perfect matchings of this
theory as follows from eq. (\ref{pmdp2}). From the height functions provided 
in the figure, we see that ${\tilde p}_1,\cdots {\tilde p}_5$ are indeed 
the internal perfect matchings of the theory. The height functions also 
reproduce the correct multiplicities in the toric diagram of the 
second phase of $dP_2$ \cite{symmtordual}.

Let us summarise the discussion in this section. Here, we have implemented 
the inverse algorithm of \cite{fhh}, starting from the matching matrix
of the parent orbifolds of the form $\BC^3/\BZ_m\times\BZ_n$, with
$(m,n) = (2,2)$, $(2,3)$ and $(3,3)$. This approach gives us a way of 
directly handling the vevs of the bifundamental fields in this theory. 
We have worked out several examples, some of which involved adjoint
fields, which we have seen how to handle in the language of
dimer models, by looking at the consistency of the matching matrix of the
daughter singularities. We have, in two examples, checked our conjecture
2 stated in section 2. We have also seen how to rule out unphysical
toric theories, apart from an apparent puzzle that we mentioned in subsection
4.2. These are the main results of this section. 

\section{Discussions}

In this paper, we have performed a detailed analysis of certain aspects
of dimer models corresponding to abelian orbifold singularities in string 
theory. To begin with, we discussed cyclic orbifolds, and studied the same 
using a combination of open and closed string techniques. In particular,
we addressed the issues of symmetries of dimer models from a closed string
perspective. Further, we have performed detailed analysis of 
Higgsing of non cyclic
orbifolds of $\BC^3$, including the simplest case where the orbifolding
action is asymmetric. We have seen how the dimer model naturally
incorporates the adjoint fields which typically arise in these cases. 
Clearly, these methods will be applicable to any abelian orbifold 
singularity, although the explicit computation of the perfect matchings 
become prohibitively difficult after the first few simple cases. 
The method of writing the perfect matchings from
the Kasteleyn matrix is helpful in these situations. Using this, 
we have verified the two conjectures that we stated in section (2)
for the orbifold $\BC^3/\BZ_3 \times \BZ_5$ although the results are
too long to present here. The methods discussed in this paper illustrate 
the general inverse procedure involving dimers, for orbifold theories. 
As a future application of these results, it would be interesting to 
understand how (a possible variant of) dimer models capture the 
combinatorics of non-supersymmetric orbifolds which have localised 
closed string tachyons in their spectrum. In those situations, 
one typically deals with the case where $\alpha'$ 
corrections are very small, i.e 
one is restriced to the classical moduli space of the gauge theory that 
lives on the world volume of the D-brane probing the orbifold. It 
can be checked that for such orbifolds \footnote{A typical example in 
the two dimensional case is the orbifold $\BC^2/\BZ_{n(p)}$ with the 
orbifolding action on the coordinates of $\BC^2$ being $\left(Z_1,Z_2\right) \to 
\left(\omega Z_1, \omega^p Z_2\right)$ with $\omega = e^{\frac{2\pi i}{n}}$.
Higher dimensional non-supersymmetric orbifolds have also been well studied
in the literature.} the main concepts of dimer models break down, and that
one needs a possible generalisation of the latter. As is known, such
non-supersymmetric orbifolds exhibit flow properties (analogous to RG
flows) to lower rank orbifolds, and it would be interesting 
to understand this dynamics from the 
standpoint of graph theory. Further, in the case of supersymmetric orbifolds,   
it would be interesting to explicitly prove our conjecture that in the 
case of generic toric varieties, the face symmetries of the corresponding 
dimer model involve only internal points in the toric diagram, whenever 
these are present.\footnote{We thank K. Kennaway for giving us a 
suggestion on how to prove this.} 

\vspace{1cm}
\begin{center}
{\bf Acknowledgments}\\
\end{center}

We would like to sincerely thank Ami Hanany and Yang-Hui He for helpful
email correspondence. We would also like to thank Ajay Singh for computer 
related help.

\newpage
\section{Appendix}

In this appendix, we provide some of the details of our calculations that have
not been provided in the main text. 

\begin{figure}[h]
\centering
\epsfxsize=4.5in
\hspace*{0in}\vspace*{.2in}
\epsffile{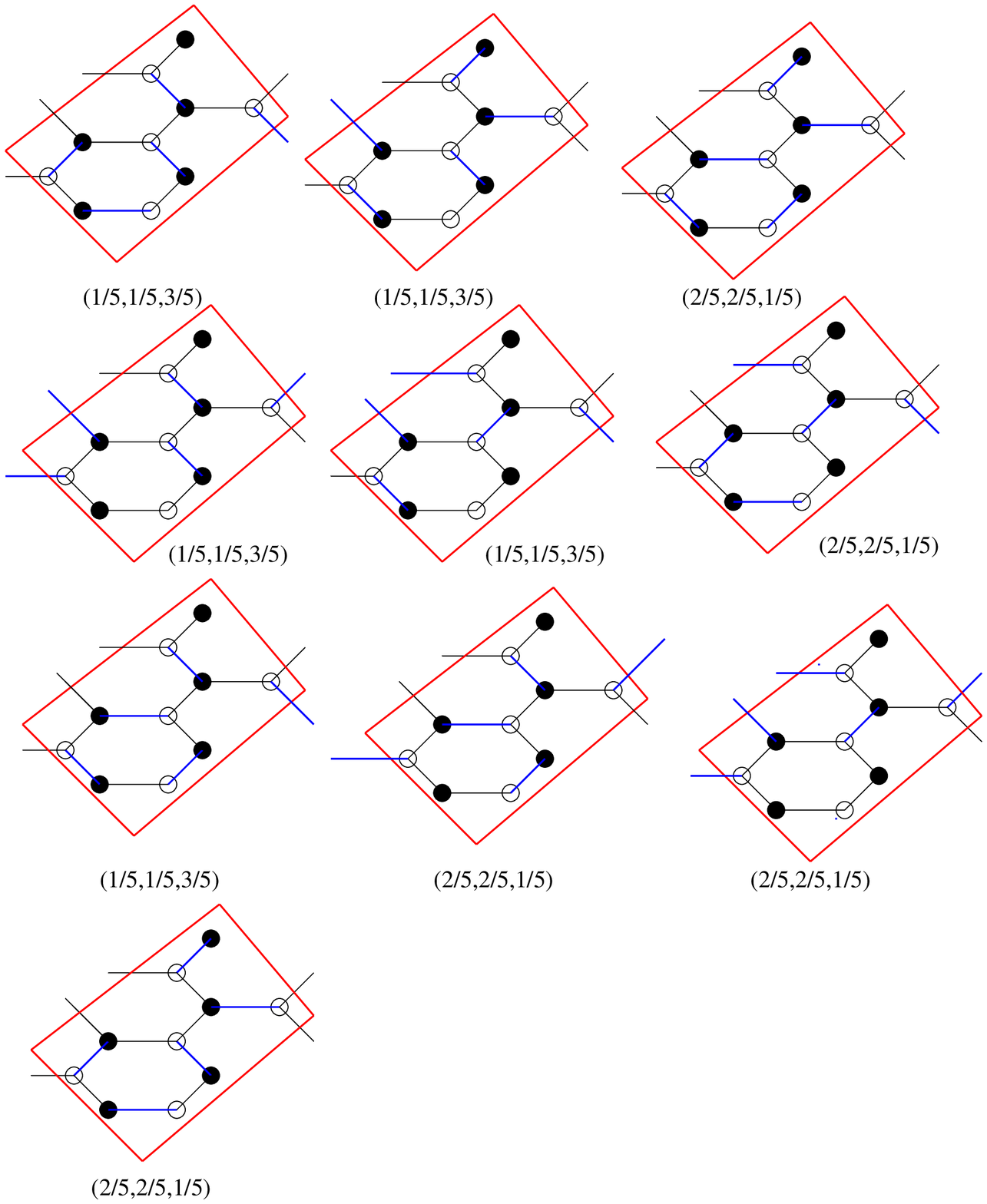}
\caption{The ten perfect matchings for the orbifold $\BC^3/\BZ_5$}
\label{app0}
\end{figure}

We start with the singularity $\BC^3/\BZ_5$. This orbifold has two twisted sectors,
as mentioned in the main text, with twisted sector R-charges 
$\left(\frac{1}{5},\frac{1}{5},\frac{3}{5}\right)$ and 
$\left(\frac{2}{5},\frac{2}{5},\frac{1}{5}\right)$. The ten perfect matchings
for this orbifold, whose dimer covering is given in fig. (\ref{c3z5}) is
shown in the fig. (\ref{app0}).  
%%%%%%%%%%%%%%%%%%%%%%%%%%%%
%%%%%%%%%%%%%%%%%%%%%%%%%%%%

\newpage
For the singularity $\BC^3/\BZ_2\times\BZ_3$, the fundamental region and  
matching matrix are shown below. The perfect matchings for this orbifold are 
shown next. For reference, we have also indicated the height functions along with
each matching.  
%%%%%%%%%%%%%%%%%%%%%%%%%%%%
\begin{figure}[h]
\centering
\epsfxsize=2.5in
\hspace*{0in}\vspace*{.2in}
\epsffile{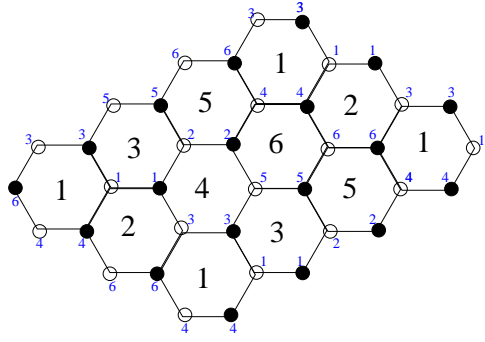}
\caption{The fundamental region for the dimer covering of the orbifold
$\BC^3/\BZ_2\times\BZ_3$. We have used slightly different labeling 
conventions compared to what appears in the main text.}
\label{app1}
\end{figure}
%%%%%%%%%%%%%%%%%%%%%%%%%%%%
{\small
\begin{eqnarray}
{\cal M}= 
\pmatrix{
~&p_1&p_2&p_3&p_4&p_5&p_6&p_7&p_8&p_9&p_{10}&
p_{11}&p_{12}&p_{13}&p_{14}&p_{15}&p_{16}&p_{17}\cr
X_1&1&0&0&1&0&1&0&1&0&1&0&0&0&0&0&1&0\cr
X_2&1&0&1&0&0&1&1&0&0&0&1&0&0&0&1&0&0\cr
X_3&1&0&0&0&1&0&1&1&0&1&0&0&0&1&0&0&0\cr
X_4&1&0&0&1&0&1&0&1&0&0&1&0&1&0&0&0&0\cr
X_5&1&0&1&0&0&1&1&0&0&1&0&1&0&0&0&0&0\cr
X_6&1&0&0&0&1&0&1&1&0&0&1&0&0&0&0&0&1\cr
Y_1&0&1&1&1&0&1&0&0&0&0&0&0&0&0&1&1&0\cr
Y_2&0&1&1&0&1&0&1&0&0&0&0&0&0&1&1&0&0\cr
Y_3&0&1&0&1&1&0&0&1&0&0&0&0&1&1&0&0&0\cr
Y_4&0&1&1&1&0&1&0&0&0&0&0&1&1&0&0&0&0\cr
Y_5&0&1&1&0&1&0&1&0&0&0&0&1&0&0&0&0&1\cr
Y_6&0&1&0&1&1&0&0&1&0&0&0&0&0&0&0&1&1\cr
Z_1&0&0&0&0&0&0&0&0&1&1&0&0&0&1&1&1&0\cr
Z_2&0&0&0&0&0&0&0&0&1&0&1&0&1&1&1&0&0\cr
Z_3&0&0&0&0&0&0&0&0&1&1&0&1&1&1&0&0&0\cr
Z_4&0&0&0&0&0&0&0&0&1&0&1&1&1&0&0&0&1\cr
Z_5&0&0&0&0&0&0&0&0&1&1&0&1&0&0&0&1&1\cr
Z_6&0&0&0&0&0&0&0&0&1&0&1&0&0&0&1&1&1\cr
}
\label{c3z2z31}
\end{eqnarray}}

\newpage
%%%%%%%%%%%%%%%%%%%%%%%%%%%%
\begin{figure}[h]
\centering
\epsfxsize=5.5in
\hspace*{0in}\vspace*{.2in}
\epsffile{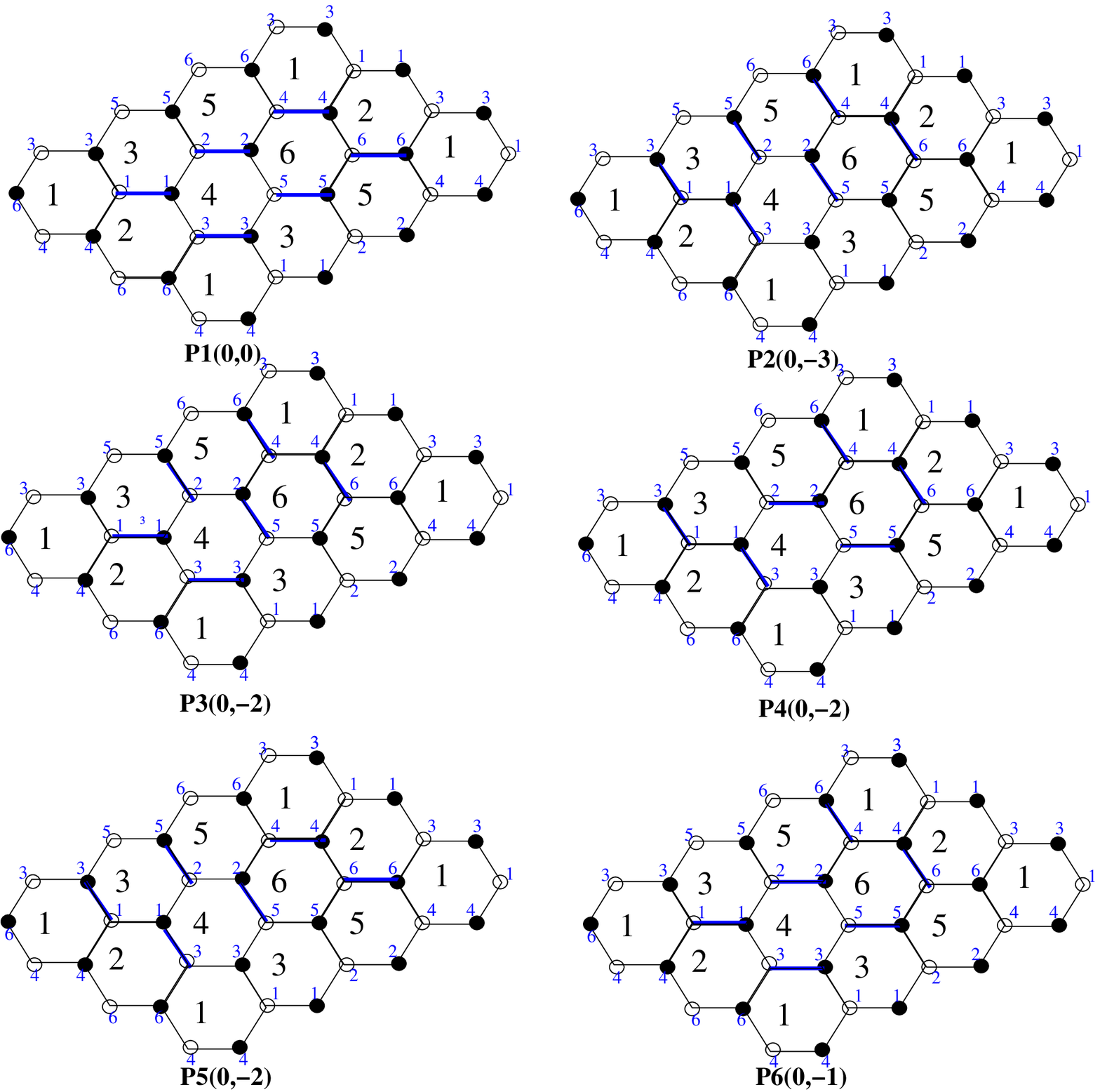}
\caption{Perfect matchings for the orbifold $\BC^3/\BZ_2\times\BZ_3$}
\label{app2}
\end{figure}
%%%%%%%%%%%%%%%%%%%%%%%%%%%%
\newpage
\begin{figure}[h]
\centering
\epsfxsize=5.5in
\hspace*{0in}\vspace*{.2in}
\epsffile{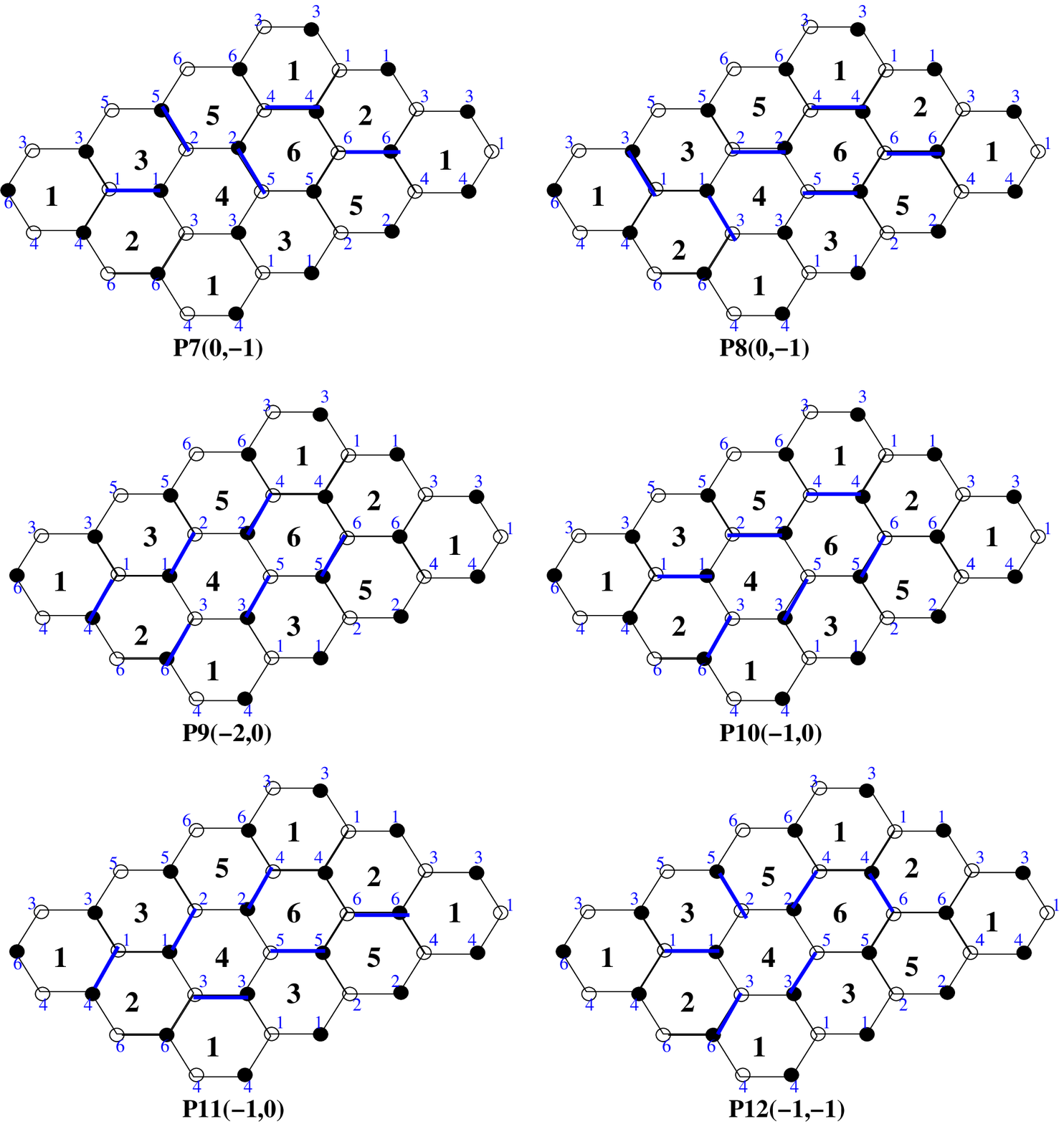}
\caption{Perfect matchings for the orbifold $\BC^3/\BZ_2\times\BZ_3$ (Contd.)}
\label{app3}
\end{figure}
%%%%%%%%%%%%%%%%%%%%%%%%%%%%
\newpage
\begin{figure}[h]
\centering
\epsfxsize=5.5in
\hspace*{0in}\vspace*{.2in}
\epsffile{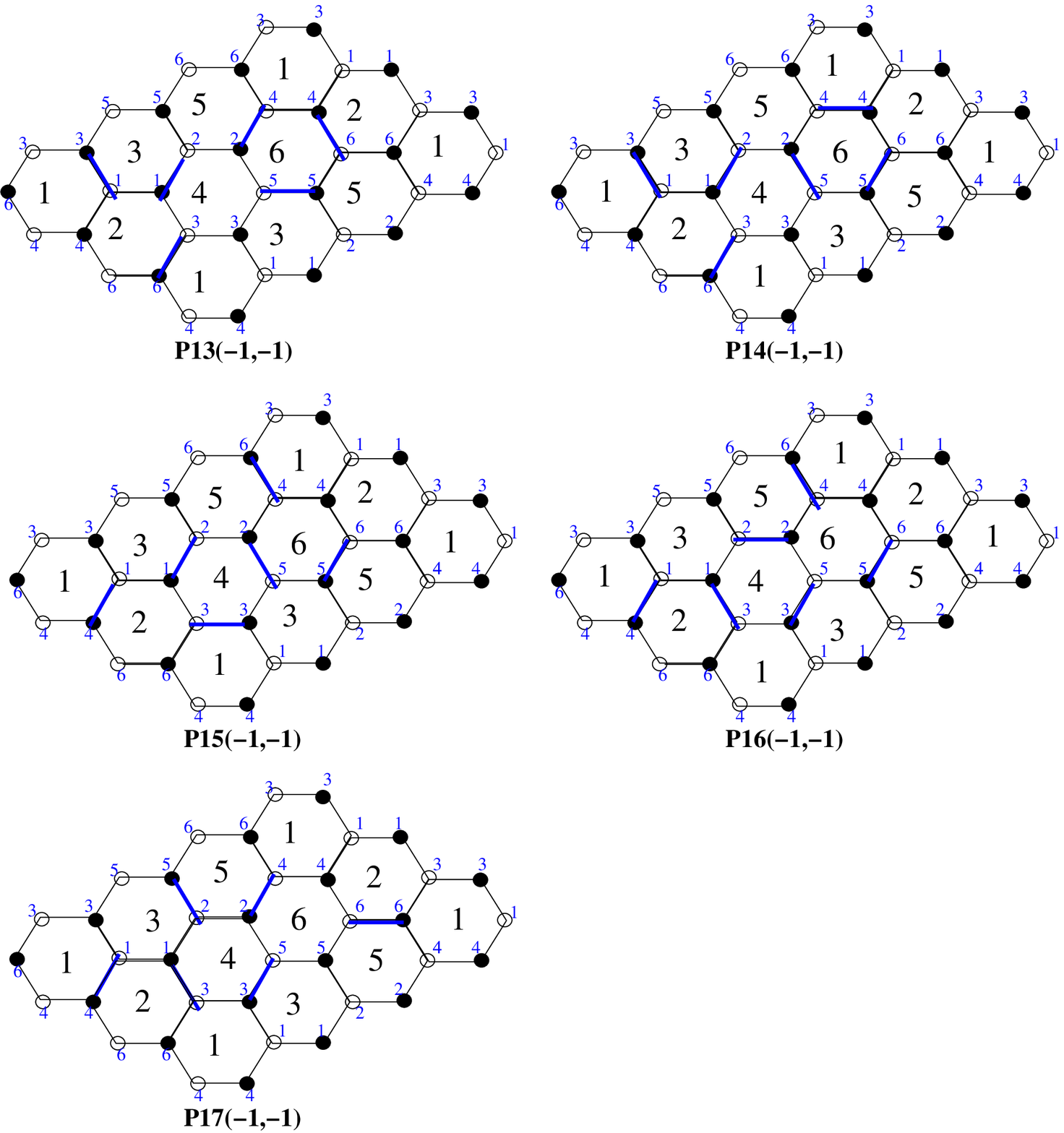}
\caption{Perfect matchings for the orbifold $\BC^3/\BZ_2\times\BZ_3$ (Contd.)}
\label{app4}
\end{figure}
%%%%%%%%%%%%%%%%%%%%%%%%%%%%
\newpage
\begin{figure}[h]
\centering
\epsfxsize=6.2in
\hspace*{0in}\vspace*{.2in}
\epsffile{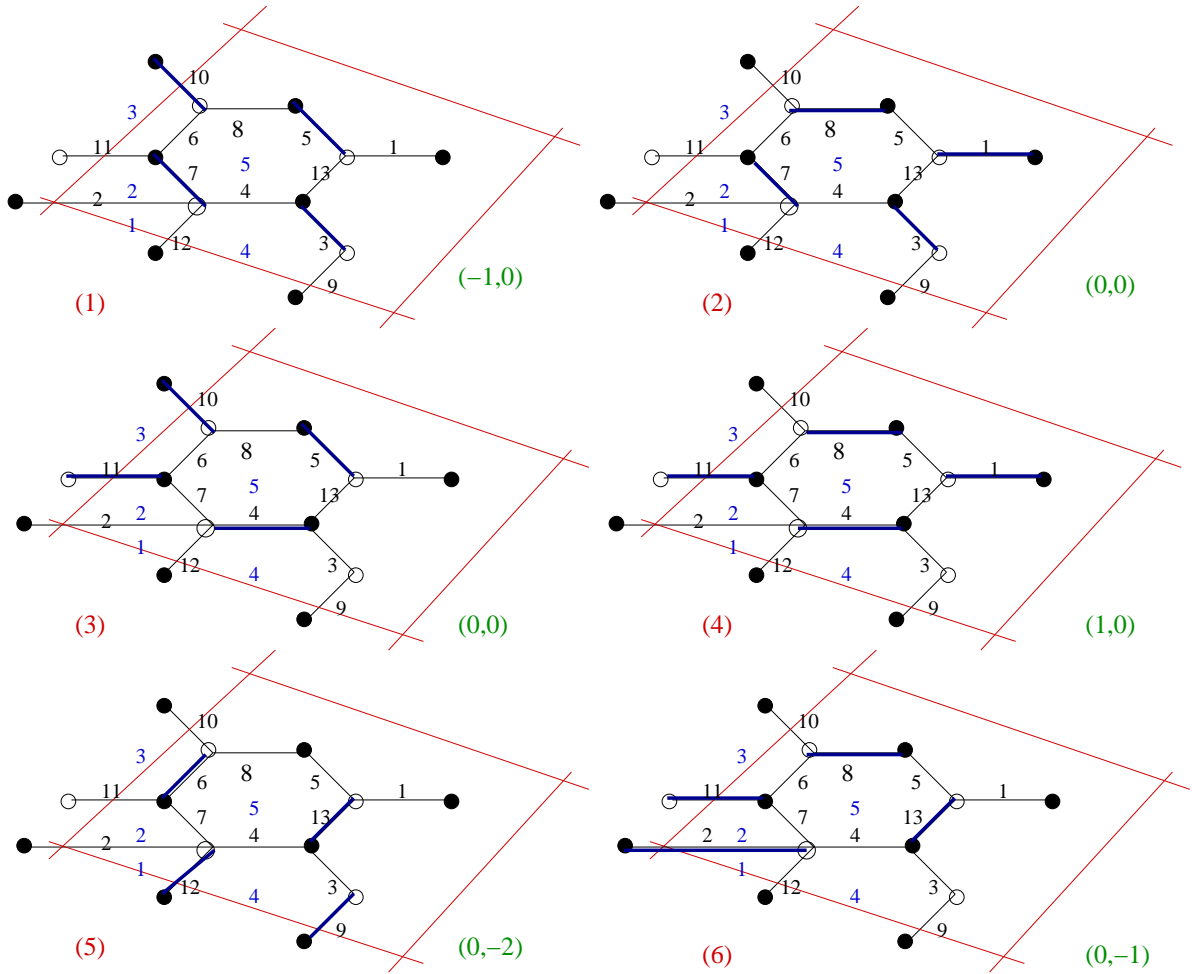}
\caption{Perfect matchings for the orbifold $\BC^3/\BZ_2\times\BZ_3$ with
the field $X_1$ of eq. (\ref{c3z2z31}) removed. The integers (in red) 
on the left of each perfect matching correspond to the matching number in the 
matrix of eq. (\ref{c3z2z3x12rem}). The height functions are indicated
by the green integers on the right.}
\label{new2a}
\end{figure}
%%%%%%%%%%%%%%%%%%%%%%%%%%%%
\newpage
\begin{figure}[h]
\centering
\epsfxsize=6.2in
\hspace*{0in}\vspace*{.2in}
\epsffile{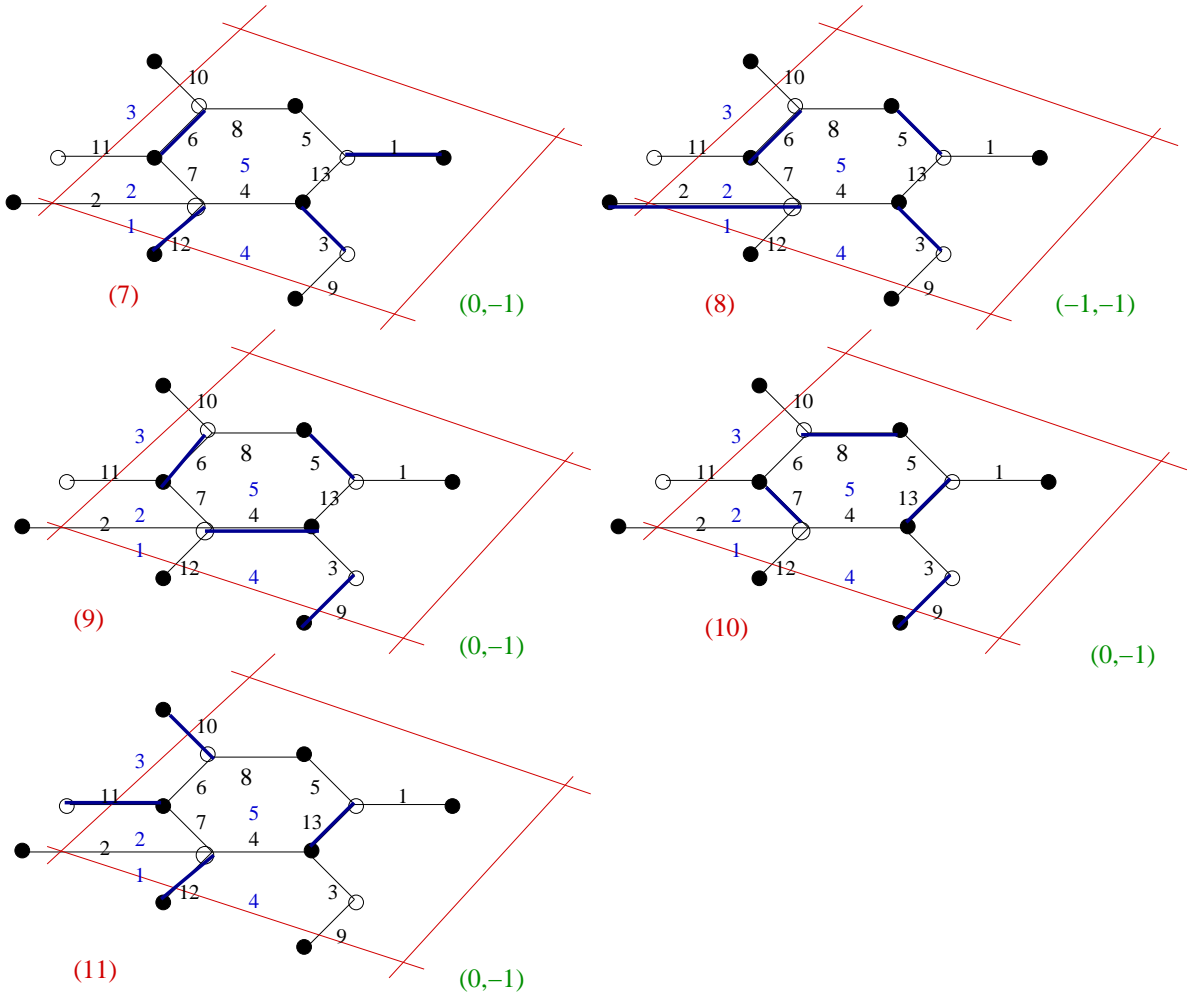}
\caption{Continued : Perfect matchings for the orbifold 
$\BC^3/\BZ_2\times\BZ_3$ with the field $X_1$ of eq. (\ref{c3z2z31}) 
removed. The same conventions as fig. (\ref{new2a}) are followed.}
\label{new2b}
\end{figure}
%%%%%%%%%%%%%%%%%%%%%%%%%%%%

\newpage
For the orbifold $\BC^3/\BZ_3\times\BZ_3$, the matching matrix is \\

$$\tiny{\left(\matrix
{0& 1& 2& 3& 4& 5& 6&7&8&9&10&11&12&13&14&15&16&17&18&19&20&21&22&23&24&25&26\cr
&&&&&&&&&&&
27&28&29&30&31&32&33&34&35&36&37&38&39&40&41&42\cr
 X1& 1& 0& 0& 0& 1& 0& 1& 1& 0& 1& 0& 0& 0& 0& 0&
    1& 1& 0& 0& 0& 0& 1& 0& 1& 1& 0\cr
&&&&&&&&&&&
 0& 0& 0& 1& 1& 0& 0& 0& 0& 0& 0& 1& 0&
    0& 1& 0\cr
    X2& 1& 0& 0& 0& 1& 0& 1& 1& 0&
   0& 0& 1& 0& 0& 0& 0& 1& 1& 0& 0& 0& 0& 0& 0& 0& 0
\cr
&&&&&&&&&&&
0& 0& 0& 0& 0& 0& 0& 0&
    0& 1& 1& 1& 1& 1& 1& 1\cr
X3& 1& 0& 0& 0& 1& 0& 1& 1& 0& 0& 1& 0& 0& 0&
    0& 1& 0& 1& 0& 0& 0& 0& 0& 0& 0& 0\cr
&&&&&&&&&&&
0& 0& 0& 1& 1& 1& 1& 0& 1& 0& 1& 0& 0& 1& 0& 0\cr
 X4& 1& 0& 1& 0& 0& 1& 1& 0& 0& 
   1& 0& 0& 0& 0& 0& 1& 1& 0& 0& 0& 0& 1& 1& 0& 0& 0\cr
&&&&&&&&&&&
0& 0& 1& 0& 0& 0& 0& 1& 0& 1& 0& 1& 1& 0& 0& 0\cr
X5& 1& 0& 1& 0& 0& 1& 1& 0& 0& 0& 0& 1& 0& 0&
0& 0& 1& 1& 0& 0& 0& 0& 0& 0& 0\cr 
&&&&&&&&&&& 1&0&1& 0& 0& 1& 1& 0& 0& 1& 1& 0&
    0& 0& 0& 0\cr
X6& 1& 0& 1& 0& 0& 1& 1& 0& 0&
   0& 1& 0& 0& 0& 0& 1& 0& 1& 0& 0& 0& 1& 1& 1& 0& 1\cr
&&&&&&&&&&&
 0& 1& 0& 1& 0& 1& 0& 0& 0& 0& 0& 0& 0& 0& 0& 0\cr
X7& 1& 0& 0& 1& 0& 1& 0& 1& 0& 1& 0& 0& 0& 0& 
0& 1& 1& 0& 0& 0& 0& 0& 0& 0& 1& 0\cr
&&&&&&&&&&&
1& 0& 1& 0& 1& 0& 1& 1& 1& 0& 0& 0& 0& 0& 0& 0\cr
X8& 1& 0& 0& 1& 0& 1& 0& 1& 0& 0& 0& 1& 0& 0& 0& 0& 1& 1& 0&
    0& 0& 0& 0& 1&1&1\cr
&&&&&&&&&&&
  1& 1& 0& 0& 0& 0& 0& 0& 0& 0& 0& 0& 0& 0& 1& 1\cr
X9& 1& 0& 0& 1& 0& 1& 0& 1& 0& 0& 1& 0& 0& 0& 0& 1& 0& 1& 0& 0& 0&
    0& 1& 0& 0& 0\cr
&&&&&&&&&&&
 0& 1& 0& 0& 0& 0& 0& 1& 1& 0& 0& 0& 1& 1& 0& 1\cr
Y1& 0& 0& 0& 0& 0& 0& 0& 0& 1& 1& 0& 1& 1& 0& 1& 0& 1& 0& 0& 1& 0& 1&
  0& 1& 1& 1\cr
&&&&&&&&&&&
 1& 0& 0& 0& 0& 0& 0& 0& 0& 0& 0& 1& 0& 0& 1& 0\cr
Y2& 0& 0& 0& 0& 0& 0& 0& 0& 1& 0& 1& 1& 1& 1& 0& 0& 0& 1& 1& 0& 0& 
0& 1& 0& 0& 0\cr
&&&&&&&&&&&
0& 1& 0& 0& 0& 0& 0& 0& 0& 1& 1& 0& 1& 1& 0& 1\cr
Y3& 0& 0& 0& 0& 0& 0& 0& 0& 1& 1& 1& 0& 0& 1& 1& 1& 0& 0& 0& 0& 1& 0&
  0& 0& 0& 0\cr
&&&&&&&&&&&
 0& 0& 1& 1& 1& 1& 1& 1& 1& 0& 0& 0& 0& 0& 0& 0\cr
Y4& 0& 0& 0& 0& 0& 0& 0& 0& 1& 1& 0& 1& 0& 1& 1& 0& 1& 0& 0& 0& 
1& 0& 0& 0& 0& 0\cr
&&&&&&&&&&&
0& 0& 1& 0& 0& 0& 0& 1& 0& 1& 0& 1& 1& 0& 1& 1\cr
Y5& 0& 0& 0& 0& 0& 0& 0& 0&
    1& 0& 1& 1& 1& 0& 1& 0& 0& 1& 0& 1& 0& 0& 0& 0& 0& 1\cr
&&&&&&&&&&&
1& 0& 0& 0& 0& 1& 1& 0& 1& 0& 1& 0& 0& 1& 0& 0\cr
Y6& 0& 0& 0& 0& 0& 0& 0& 0& 1& 1& 1& 0& 1& 1& 0& 1& 0& 0& 1& 0& 0& 1& 1& 
1& 1& 0 \cr
&&&&&&&&&&&
 0& 1& 0& 1& 1& 0& 0& 0& 0& 0& 0& 0& 0& 0& 0& 0\cr
Y7& 0& 0& 0& 0& 0& 0& 0& 0& 1& 
     1& 0& 1& 1& 1& 0& 0& 1& 0& 1& 0& 0& 0& 0& 0& 1& 0\cr
&&&&&&&&&&&
1& 0& 1& 0& 1& 0& 1& 0& 0& 1& 1& 0& 0& 0& 0& 0\cr
Y8& 0& 0& 0& 0& 0& 0& 0& 0& 1& 0& 1& 1& 0& 1& 1& 0& 0& 1& 0& 0& 1& 0& 0& 1& 
0& 1\cr
&&&&&&&&&&&
0&1& 0& 1& 0& 1& 0& 0& 0& 0& 0& 0& 0& 0& 1& 1\cr
Y9& 0& 0& 0& 0& 0& 0& 0& 0& 1& 1& 1& 0& 1& 0& 1& 1& 0&
     0& 0& 1& 0& 1& 1& 0& 0& 0\cr 
&&&&&&&&&&&
 0& 0& 0& 0& 0& 0& 0& 1& 1& 0& 0& 1& 1& 1& 0& 0\cr 
Z1& 0& 1& 1& 0& 1& 0& 1& 0& 0& 0& 0& 0& 0& 0& 1& 0& 0& 0& 0& 1& 1& 1& 0& 1& 
0& 1\cr
&&&&&&&&&&&
0& 0& 0& 1& 0& 1& 0& 0& 0& 0&0& 1& 0& 0& 1& 0\cr
Z2& 0& 1& 1& 0& 1& 0& 1& 0& 0& 0& 0& 0& 1& 0& 0& 0& 0& 0& 1& 1& 0& 1& 1& 0& 
0& 0\cr
&&&&&&&&&&&
0& 0& 0& 0& 0& 0& 0& 0& 0& 1& 1& 1& 1& 1& 0& 0\cr
Z3& 0& 1& 1& 0& 1& 0& 1& 0& 0& 0& 0& 0& 0& 1& 0& 0& 0& 0& 1& 0& 1& 0& 0& 0&
 0& 0\cr
&&&&&&&&&&&
0& 0& 1& 1& 1& 1& 1& 0& 0& 1& 1& 0& 0& 0& 0& 0\cr
Z4& 0& 1& 1& 1& 0& 1& 0& 0& 0& 0& 0& 0& 0& 1& 0& 0& 0& 0& 1& 0& 1& 0& 1&
    0& 0& 0\cr
&&&&&&&&&&&
0& 1& 1& 0& 0& 0& 0& 1& 0& 1& 0& 0& 1& 0& 0& 1\cr
Z5& 0& 1& 1& 1& 0& 1& 0& 0& 0& 0& 0& 0& 0& 0& 1& 0& 0& 0& 0& 1& 1& 0& 0& 0& 
0& 1\cr
&&&&&&&&&&&
1& 0& 1& 0& 0& 1& 1& 1& 1& 0& 0& 0& 0& 0& 0& 0 \cr
Z6& 0& 1& 1& 1& 0& 1& 0& 0& 0& 0& 0& 0& 1& 0& 0& 0& 0& 0& 1& 1& 0& 1& 1& 1& 
1& 1\cr
&&&&&&&&&&&
1& 1& 0& 0& 0& 0& 0& 0& 0& 0& 0& 0& 0& 0& 0& 0\cr
Z7& 0& 1& 0& 1& 1& 0& 0& 1& 0& 0& 0& 0& 1& 0& 0& 0& 0& 0& 1& 1& 0& 0& 0& 0& 
1& 0\cr
&&&&&&&&&&&
1& 0& 0& 0& 1& 0& 1& 0& 1& 0& 1& 0& 0& 1& 0& 0\cr
Z8& 0& 1& 0& 1& 1& 0& 0& 1& 0& 0& 0& 0& 0& 1& 0& 0& 0& 0& 1& 0& 1& 0& 0& 1&
 1& 0\cr
&&&&&&&&&&&
0& 1& 0& 1& 1& 0& 0& 0& 0& 0& 0& 0& 0& 0& 1& 1\cr
Z9& 0& 1& 0& 1& 1& 0& 0& 1& 0& 0& 0& 0& 0& 0& 1& 0& 0& 0& 0& 1& 1& 0& 0& 0& 
0& 0\cr
&&&&&&&&&&&
0& 0& 0& 0& 0& 0& 0& 1& 1& 0& 0& 1& 1& 1& 1& 1
}\right)} \label{z3z3pm}$$

\begin{figure}[h]
\centering
\epsfxsize=5.5in
\hspace*{0in}\vspace*{.2in}
\epsffile{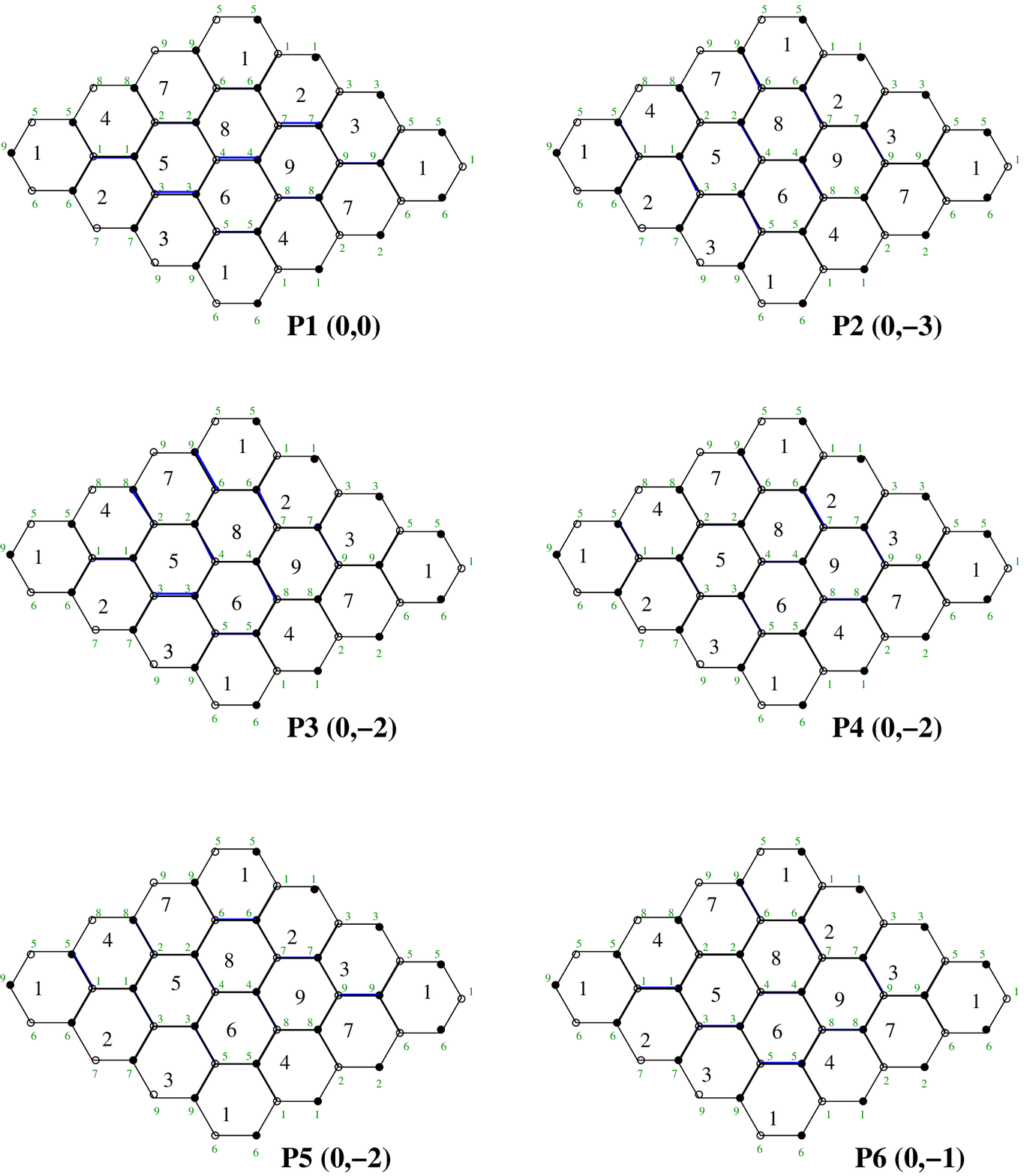}
\caption{Some of the perfect matchings corresponding to the points on the
edges of the toric diagram for the orbifold 
$\BC^3/\BZ_3\times\BZ_3$. The corresponding height functions are given 
for reference. The closed string twisted sector R-charges can be read off
by assigning weights $\left(\frac{1}{9},0,0\right)$, 
$\left(0,\frac{1}{9},0\right)$ and $\left(0,0,\frac{1}{9}\right)$ to 
the three types of edges.} 
\label{app5}
\end{figure}
%%%%%%%%%%%%%%%%%%%%%%%%%%%%
\newpage
\begin{figure}[h]
\centering
\epsfxsize=5.5in
\hspace*{0in}\vspace*{.2in}
\epsffile{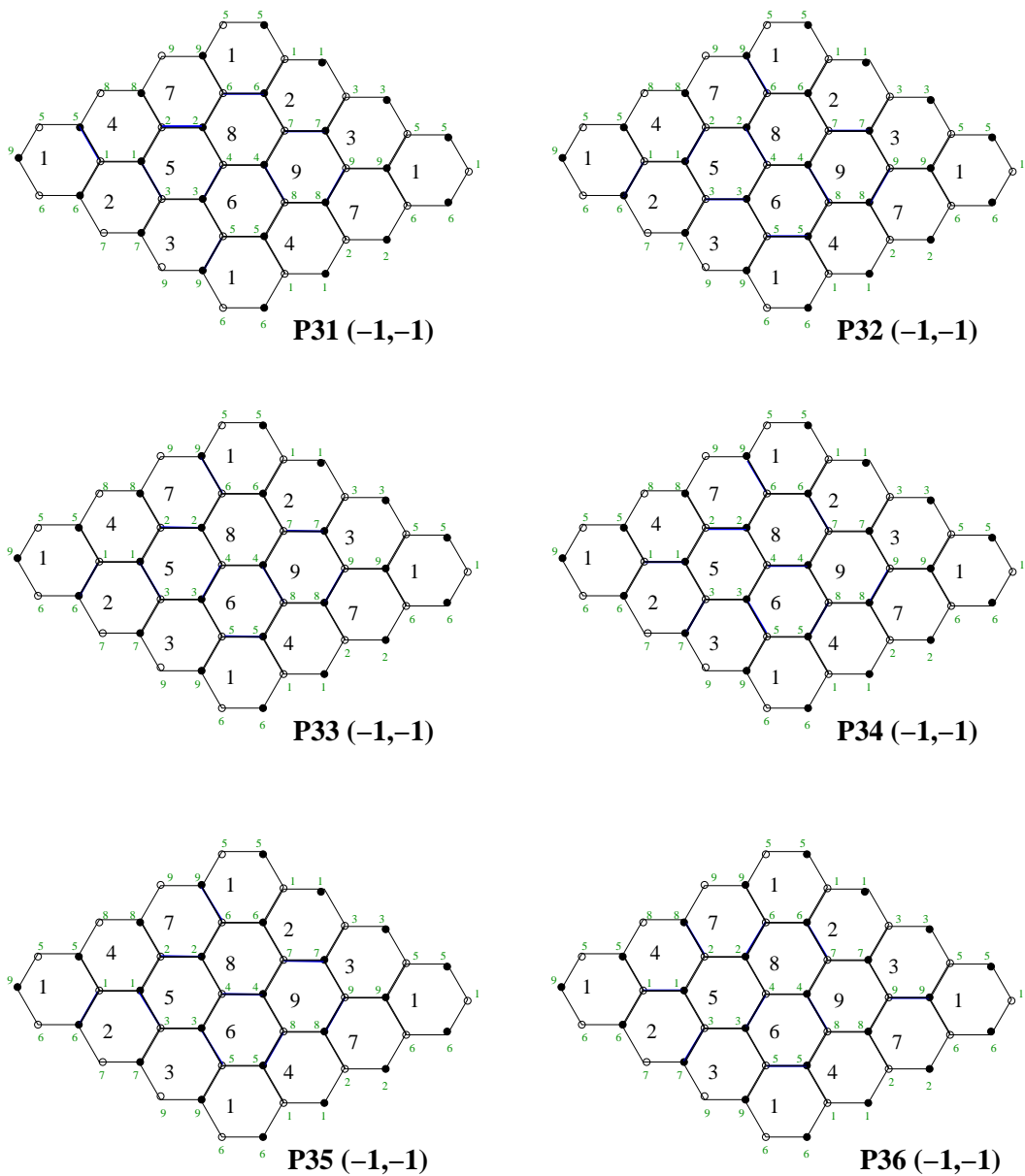}
\caption{Some perfect matchings corresponding to the internal points in
the toric diagram of the orbifold $\BC^3/\BZ_3\times\BZ_3$}
\label{app6}
\end{figure}
%%%%%%%%%%%%%%%%%%%%%%%%%%%%
\newpage
\begin{figure}[h]
\centering
\epsfxsize=6.5in
\hspace*{0in}\vspace*{.2in}
\epsffile{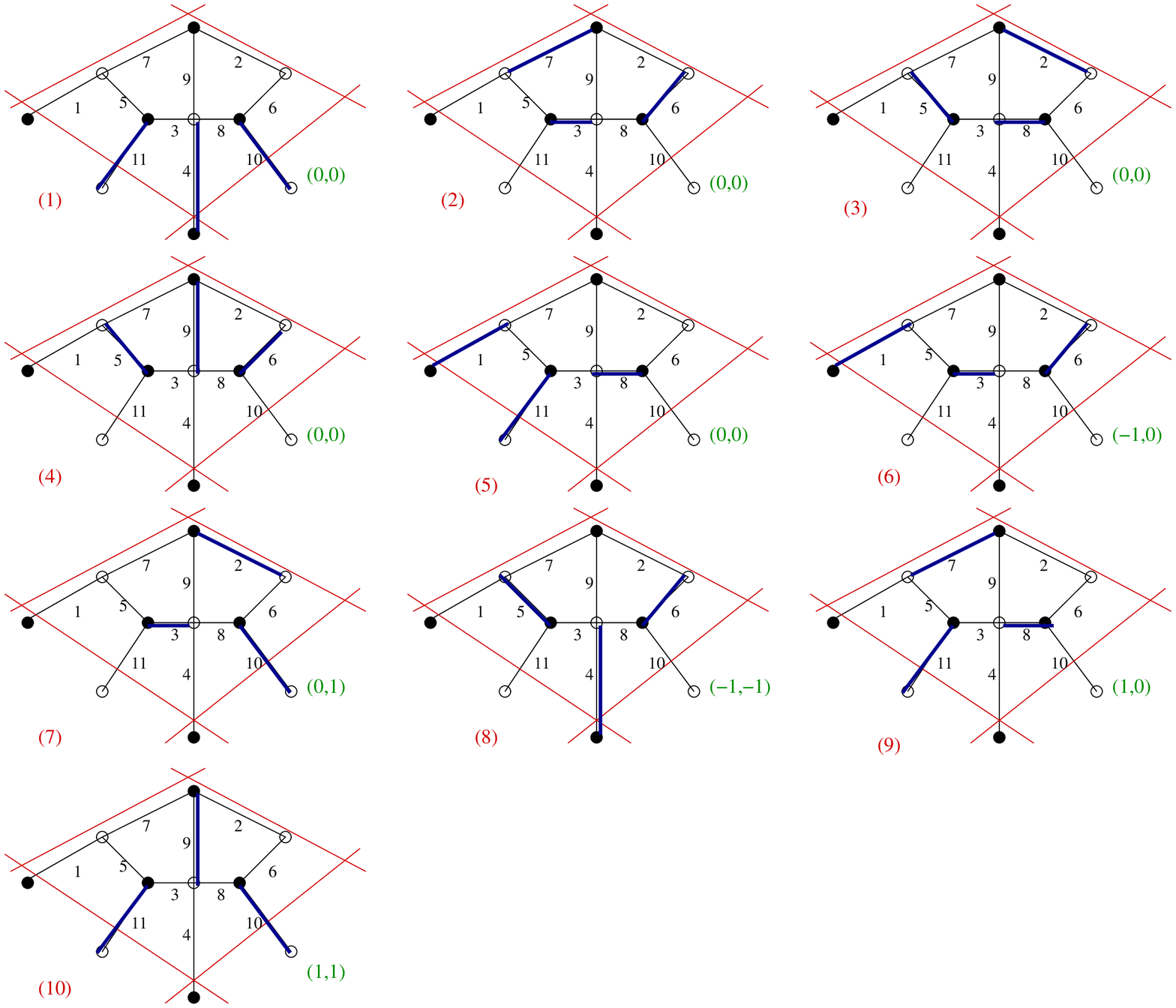}
\caption{Perfect matchings corresponding to the model II of the $dP_2$
singularity, drawn from eq. (\ref{pmdp2}). 
The perfect matching numbers are denoted by the red integers,
while the green integers denote the height functions.}
\label{dp2pm}
\end{figure}
%%%%%%%%%%%%%%%%%%%%%%%%%%%%


\begin{thebibliography}{9}
\bibitem{clpp}
M. Cvetic, H. Lu, D. N. Page, C. N. Pope, ``New Einstein-Sasaki spaces in 
five and higher dimensions,'' Phys. Rev. Lett. {\bf 95} 071101 (2005),
{\tt hep-th/0504225}
\bibitem{ms}
D. Martelli, J. Sparks, ``Toric Sasaki-Einstein metrics on $S^2$ and
$S^3$,'' Phys. Lett. {\bf B 621}, 208 (2005), {\tt hep-th/0505027}
\bibitem{dm}
M. R. Douglas, G. W. Moore, ``D-branes, quivers, and ALE instantons,''
{\tt hep-th/9603167}
\bibitem{dgm}
M. R. Douglas, B. R. Greene, D. R. Morrison, 
``Orbifold resolution by D-branes,'' Nucl. Phys. {\bf B506} (1997) 84,
{\tt hep-th/9704151}
\bibitem{ms1}
D. Martelli, J. Sparks,
``Toric geometry, Sasaki-Einstein manifolds and a new infinite class of
AdS/CFT duals,''
Comm. Math. Phys. {\bf 262} (2006) 51, {\tt hep-th/0411238}
\bibitem{bfhms}
S. Benvenuti, S. Franco, A. Hanany, D. Martelli, J. Sparks,
``An Infinite family of superconformal quiver gauge theories with
Sasaki-Einstein duals,''
JHEP {\bf 0506} (2005) 064, {\tt hep-th/0411264}
\bibitem{kast}
P. Kasteleyn, ``Graph theory and crystal physics,'' in {\tt
Graph theory and theoretical physics}, pp 43 - 110, Academic Press,
London, 1967.
\bibitem{kenyon}
R. Kenyon, ``An introduction to the dimer model,'' {\tt
math.CO/0310236}
\bibitem{han1}
A. Hanany, K. D. Kennaway, ``Dimer models and toric diagrams,'' 
{\tt hep-th/0503149}
\bibitem{fhkvw}
S. Franco, A. Hanany, K. D. Kennaway, D. Vegh, B. Wecht,
``Brane dimers and quiver gauge theories,''
JHEP {\bf 0601} 2006, 096, {\tt hep-th/0504110}
\bibitem{kennaway}
K. D. Kennaway, ``Brane Tilings,''
Int. J. Mod. Phys. {\bf A22} 2007, 2977, {\tt arXiv:0706.1660 [hep-th]}
\bibitem{yamazaki}
M. Yamazaki, ``Brane Tilings and Their Applications,''
{\tt arXiv:0803.4474 [hep-th]}
\bibitem{masterspace}
D. Forcella, A. Hanany, Y-H. He, A. Zaffaroni, 
``The Master Space of N=1 Gauge Theories,'' 
{\tt arXiv:0801.1585 [hep-th]}
\bibitem{tapodimer}
T. Sarkar, ``On Dimer Models and Closed String Theories,''
JHEP {\bf 0710} 2007 010, {\tt arXiv:0705.3575 [hep-th]}
\bibitem{fhh}
B. Feng, A. Hanany, Y-H He, 
`` D-brane gauge theories from toric singularities and toric duality,''
Nucl. Phys. {\bf B595} (2001) 165, {\tt hep-th/0003085}
\bibitem{bglp}
C. Beasley, B. R. Greene, C.I. Lazaroiu, M. R. Plesser,
`` D3-branes on partial resolutions of Abelian quotient singularities of 
Calabi-Yau threefolds,'' Nucl. Phys. {\bf B566} 2000, 599,
{\tt hep-th/9907186}
\bibitem{mp}
D. R. Morrison, M. R. Plesser,
``Nonspherical Horizons 1,'' Adv. Theor. Math. Phys. {\bf 3}, 1999, 1,
{\tt hep-th/9810201}
\bibitem{francovegh}
S. Franco, D. Vegh, ``Moduli spaces of gauge theories from dimer models : 
Proof of the correspondence,'' JHEP {\bf 0611} (2006) 054, 
{\tt hep-th/0601063}
\bibitem{pru}
J. Park, R. Rabadan, A. M. Uranga, ``Orientifolding the conifold,''
Nucl. Phys. {\bf B570} 2000 38, {\tt hep-th/9907086}
\bibitem{tapo2}
T. Sarkar, `` On localized tachyon condensation in $\BC^2 / \BZ_n$
and $\BC^3/\BZ_n$,'' Nucl. Phys. {\bf B700} (2004) 490,
{\tt hep-th/0407070}
\bibitem{wittenphases}
E. Witten, ``Phases of N=2 Theories in Two Dimensions,''
Nuclear Physics {\bf B 403}, (1993) 159, {\tt hep-th/9301042}.
\bibitem{uranga2}
I. Garcia-Etxebarria, F. Saad, A. M. Uranga,
``Quiver gauge theories at resolved and deformed singularities using dimers,''
JHEP {\bf 0606} (2006) 055, {\tt hep-th/0603108}
\bibitem{muto}
T. Muto, ``D-branes on orbifolds and topology change,''
Nucl. Phys. {\bf B521} (1998) 183, {\tt hep-th/9711090}
\bibitem{tapo3}
T. Sarkar, `` D-brane gauge theories from toric singularities of the form 
$\BC^3/\Gamma$ and $\BC^4/\Gamma$,''
Nucl. Phys. {\bf B595} (2001) 201, {\tt hep-th/0005166}
\bibitem{unhigdp}
B. Feng, S. Franco, A. Hanany, Y-H. He, ``UnHiggsing the del Pezzo,''
JHEP {\bf 0308} (2003) 058, {\tt hep-th/0209228}
\bibitem{symmtordual}
B. Feng, S. Franco, A. Hanany, Y-H. He, ``Symmetries of toric duality,''
JHEP {\bf 076} (2002) 0212, {\tt hep-th/0205144}
\end{thebibliography}
\end{document}